\documentclass[sigconf]{acmart}

\usepackage{multirow}
\usepackage{array}
\usepackage{hyperref}
\usepackage{url}
\usepackage{booktabs}
\usepackage{graphicx}
\usepackage{color-edits}
\usepackage{booktabs, multirow}
\usepackage{pdflscape}
\usepackage{colortbl}
\usepackage{amsmath, amsfonts}
\usepackage{bbm}
\usepackage{nicefrac}       % compact symbols for 1/2, etc.
\usepackage{microtype}      % microtypography
\usepackage{framed}

\AtBeginDocument{%
  }

\copyrightyear{2025}
\acmYear{2025}
\setcopyright{cc}
\setcctype{by}
\acmConference[DIS '25]{Designing Interactive Systems Conference}{July 5--9, 2025}{Funchal, Portugal}
\acmBooktitle{Designing Interactive Systems Conference (DIS '25), July 5--9, 2025, Funchal, Portugal}\acmDOI{10.1145/3715336.3735745}
\acmISBN{979-8-4007-1485-6/2025/07}

\begin{document}

\renewcommand{\labelenumi}{\arabic{enumi}}

%=================================================

\title[]{Making the Right Thing: Bridging HCI and Responsible AI in Early-Stage AI Concept Selection}

%=================================================

\author{Ji-Youn Jung}
\affiliation{%
  %\institution{Human-Computer Interaction Institute}
  \institution{Carnegie Mellon University}
  \city{Pittsburgh}
  \state{PA}
  \country{USA}
}
\email{jiyounj@andrew.cmu.edu}

\author{Devansh Saxena}
\authornote{Author conducted this research as a Presidential Postdoctoral Fellow at Carnegie Mellon University.}
\affiliation{%
  \institution{University of Wisconsin-Madison}
  \city{Madison, WI}
  \country{USA}}
\email{devansh.saxena@wisc.edu}

\author{Minjung Park}
\affiliation{%
  %\institution{Human-Computer Interaction Institute}
  \institution{Carnegie Mellon University}
  \city{Pittsburgh, PA}
  \country{USA}}
\email{mpark2@andrew.cmu.edu}

\author{Jini Kim}
\affiliation{%
  %\institution{Human-Computer Interaction Institute}
  \institution{Carnegie Mellon University}
  \city{Pittsburgh, PA}
  \country{USA}}
\email{jinik@andrew.cmu.edu}

\author{Jodi Forlizzi}
\affiliation{%
  %\institution{Human-Computer Interaction Institute}
  \institution{Carnegie Mellon University}
  \city{Pittsburgh, PA}
  \country{USA}}
\email{forlizzi@cs.cmu.edu}

\author{Kenneth Holstein}
\affiliation{%
  %\institution{Human-Computer Interaction Institute}
  \institution{Carnegie Mellon University}
  \city{Pittsburgh, PA}
  \country{USA}}
\email{kjholste@andrew.cmu.edu}

\author{John Zimmerman}
\affiliation{%
  %\institution{Human-Computer Interaction Institute}
  \institution{Carnegie Mellon University}
  \city{Pittsburgh, PA}
  \country{USA}}
\email{johnz@andrew.cmu.edu}

\renewcommand{\shortauthors}{Jung et al.}

%=================================================

\begin{abstract}
AI projects often fail due to financial, technical, ethical, or user acceptance challenges—failures frequently rooted in early-stage decisions. While HCI and Responsible AI (RAI) research emphasize this, practical approaches for identifying promising concepts early remain limited. Drawing on Research through Design, this paper investigates how early-stage AI concept sorting in commercial settings can reflect RAI principles. Through three design experiments—including a probe study with industry practitioners—we explored methods for evaluating risks and benefits using multidisciplinary collaboration. Participants demonstrated strong receptivity to addressing RAI concerns early in the process and effectively identified low-risk, high-benefit AI concepts. Our findings highlight the potential of a design-led approach to embed ethical and service design thinking at the front end of AI innovation. By examining how practitioners reason about AI concepts, our study invites HCI and RAI communities to see early-stage innovation as a critical space for engaging ethical and commercial considerations together.

\end{abstract}

%=================================================




% \received{20 February 2007}
% \received[revised]{12 March 2009}
% \received[accepted]{5 June 2009}

\maketitle

\begingroup
\renewcommand\thefootnote{}\footnotetext{%
© 2025 The Author(s). This is the author’s version of the work.
It is posted here for your personal use. Not for redistribution.
The definitive Version of Record appears in \emph{Proceedings of
DIS 2025}. \url{https://doi.org/10.1145/3715336.3735745}}
\endgroup

%=================================================
%=================================================

\section{Introduction}

Artificial intelligence (AI) offers immense potential to revolutionize industries and address global challenges. It creates value by automating work and increasing efficiency (e.g., warehouse and manufacturing robots, chatbots, document summarization); forecasting and providing critical insights (e.g., financial forecasting, smart inventory management, predictive maintenance); and accelerating breakthroughs in areas that have inhumanly large problem spaces like drug discovery~\cite{mak2024artificial, gentile2022artificial} and the creation of new materials for things like batteries and computer chips~\cite{lv2022machine, pyzer2022accelerating}. The hype, hope, and promise surrounding AI have sparked widespread interest and huge investments. A recent survey showed that 40\% of CEOs plan large AI investments, and they expect these to pay off with improved competitive advantage~\cite{mcdade_more_2024}. Hundreds of billions of dollars are being spent each year with the hope that AI produces rich rewards~\cite{kindig_ai_2024, noauthor_will_2024}. 

The success and investment surrounding AI products and services imply an effective innovation process; however, recent human-computer interaction (HCI) research exposed many challenges. Currently, about 85\% of AI initiatives fail~\cite{ermakova_beyond_2021, weiner_why_2020, kidd_why_2018, joshi_why_2021} because many systems: (i) cannot generate benefits that outweigh their development and operational costs, (ii) cannot build a model that achieves the minimal required level of performance to create real value, (iii) don’t address real needs, so users won’t adopt and use them, and/or (iv) face significant ethical issues such as biased data that cause unacceptable, unintended harm. In addition to the high failure rate, researchers note a lot of missed, low-hanging fruit~\cite{yang_planning_2016, yang_re-examining_2020, yildirim_how_2022, yildirim_creating_2023}. Organizations seem to ignore situations where simple AI implementations with moderate performance would be immediately useful for users and generate value for service providers. The hype surrounding AI seems to pressure organizations to rush into AI innovation, and to take big risks instead of a more measured and steady approach. 

The Responsible AI (RAI) community has made significant pro-gress in identifying sources of unintended harm and creating a variety of practical toolkits and frameworks. However, implementing these insights within organizations often faces resistance, driven by concerns about their perceived incompatibility with existing workflows (e.g.,~\cite{deng_investigating_2023, holstein_improving_2019, lee_landscape_2021, widder_its_2023}) or fear of potentially hindering innovation~\cite{spisak_13_2023}. To address these challenges, prior research has assessed the state of adoption and explored ways to bridge gaps. For instance, Rakova et al. highlighted practitioners' aspirations to align RAI efforts with their organization's mission and values~\cite{rakova_where_2021}. Despite these efforts, a critical knowledge gap remains: how to operationalize these aspirations and effectively integrate commercial objectives with RAI principles.

A growing body of RAI researchers found that trying to fix ethical issues after models have been trained and deployed is often infeasible or even impossible in practice~\cite{cooper_emergent_2021, dutta_is_2020, kallus_residual_2018}. This has led to a self-critique within the RAI community: while adept at identifying problems, the solutions often come too late in the development process. Researchers suggest that one way to avoid RAI challenges is to address concerns during the earliest stages of the innovation process~\cite{saxena2025ai, coston_validity_2023, kawakami_situate_2024, passi_problem_2019, raji_fallacy_2022, selbst_fairness_2019, wang_against_2024, yildirim_creating_2023}. At the same time, HCI researchers suggest that improving ideation (the generation of many innovation concepts) might reduce the high AI failure rate by surfacing low-risk, high-value concepts. Recent work offers new ways to generate ideas that more effectively surface the harmonious intersection of what users need and what AI can reasonably produce~\cite{yildirim_creating_2023, liu_human-centered_2024}. Despite the shared interest and a common vision from both RAI and HCI communities in improving early-stage AI innovation, little existing work explicitly bridges these domains. Specifically, there has been little exploration of RAI principles shaping ideation processes or AI innovation research adopting RAI insights.

Our research builds on insights from RAI advocating for the early consideration of ethical harms in innovation, and from HCI research focused on improving the ideation of AI concepts. We specifically focus on concept selection, and how innovation teams could assess concepts generated during brainstorming to surface low-risk, high-benefit opportunities. HCI currently does not offer a formal process for selecting the “best” concept. We inferred that scaffolding the process might help, given the complexity of simultaneously addressing technical, financial, user-acceptance, and ethical risks. We ask the following questions: \textit{Can RAI concerns be integrated into a commercial AI innovation process at the point of project selection?} \textit{Are AI innovators open to a more formal process for assessing and selecting low-risk, high-benefit opportunities?}

Following a Research through Design approach~\cite{zimmerman_research_2007}, we conducted three design experiments including a probe study with professional AI innovators. The probe explored a more structured approach to collaboratively assessing concepts and selecting what to innovate. We found that professionals were open to integrating RAI concerns into a commercial AI innovation process, and they could leverage the collective intelligence in their team to reach a consensus. Their reflections on current practices indicated they don’t often brainstorm new product or service concepts, raising questions about who is doing this work.   

Our paper makes four contributions. First, we present a probe study on early-stage AI innovation practices. Second, we advance our understanding of how to support more effective and responsible AI innovation in practice. Third, we offer insights with the goal of improving early-stage ideation and project selection. Finally, we pose new research questions for the HCI and RAI communities, advancing the discourse on AI innovation.

\section{Related Work}
Our work draws from and attempts to integrate two separate threads of research. We build on HCI research focused on lowering the risk of failure by improving the ideation of AI concepts. We also draw on RAI research and its intention of reducing AI’s unintended harms. We are committed to RAI's goals of maximizing AI’s benefits while minimizing its harms~\cite{peters_responsible_2020}.

\subsection{HCI Research on Improving AI Innovation}
HCI research has a long history of discussing the interplay between \textit{sketching} (ideation: the envisioning of many different things to make) and \textit{prototyping} (iteration: the use of rapid prototyping to iteratively refine a system into being). Buxton, in his seminal work on user experience, discusses this as the difference between “making the \textbf{right thing}” (sketching) and “making the \textbf{thing right}” (prototyping)~\cite{buxton_sketching_2007}. Recent HCI research on AI innovation suggests that the high failure rate likely stems from poor quality ideation~\cite{yang_re-examining_2020, yang_sketching_2019, yildirim_sketching_2024}. Researchers note that data science teams often envision concepts customers don’t want, while design teams envision concepts that cannot be built. 

This body of research details several potential causes of poor-quality ideation. In many cases, HCI/UX is not invited to join projects until after deciding what to make has happened~\cite{dove_ux_2017}. This is sort of like traveling back in time to the 1990s when designers were invited to join a project late in the process. They dismissively described the goal of slapping a “pretty” interface onto an application no one wants as “putting lipstick on a pig”~\cite{cooper_inmates_1999, cooper_about_2014}. However, including HCI/UX from the start won’t solve the failure problem. Many HCI/UX practitioners find it challenging to grasp AI's capabilities — what exactly can AI do~\cite{dove_ux_2017}? The few “effective” HCI/UX practitioners employ internalized AI capability abstractions, allowing them to recognize situations when a capability might be valuable~\cite{yang_investigating_2018}. They also use a set of examples associated with these capabilities to communicate their ideas to other HCI/UX practitioners or AI collaborators. One challenge unique to AI is its high level of uncertainty. When ideating, AI innovators often struggle with understanding whether they can technically deliver a desired capability and the potential errors a system might make~\cite{yang_re-examining_2020}. As a system is deployed and accumulates data, these factors can change, influencing the system's performance. Innovators also struggle to recognize AI’s development and operational costs --- is a concept expensive or cheap~\cite{yildirim_how_2022}? Collectively, these issues all present challenges to effective ideation and project selection.

Data science plays a key role in AI innovation; however, it lacks a strong connection to human-centered design processes like ideation. Today, most data science textbooks describe the innovation process as starting with problem formulation: defining a problem, identifying constraints and resources, and scoping a project that should create value~\cite{witten_data_2002}. Data science students are not typically taught to ideate, to think of a hundred different things they might create, and then assess their collection to find the best concept. HCI research exploring the emerging role of data science in the enterprise detailed several challenges that can impact the ideation and selection of viable, valuable projects. Data scientists shared that they struggle to connect business strategy, user needs, and AI opportunities~\cite{kross_orienting_2021, nahar_collaboration_2022}. They also struggle to communicate the meaning of a model’s performance, negatively impacting collaboration with non-data science stakeholders~\cite{zhang_how_2020}. This challenge to define and communicate what is “good enough” creates tension around assessing how good an innovation concept might be. 

In support of human-centered AI innovation, HCI researchers have developed resources including design guidelines and patterns~\cite{amershi_guidelines_2019, apple_human_2023, google_people_2019}. Currently, these resources only support the prototyping phase~\cite{yildirim_investigating_2023}. Similarly, tools for machine learning practitioners often focus too narrowly on low-level technical details, neglecting the broader, conceptual exploration necessary during sketching~\cite{lam_model_2023}. Research on how HCI/UX teams employ these guidelines surfaced practitioners’ requests for additional tools and resources that can help with ideation and selection~\cite{yildirim_investigating_2023}. Research also shows that HCI/UX practitioners have started creating their own resources by documenting AI capabilities in support of ideation~\cite{yildirim_how_2022, yildirim_creating_2023}. HCI researchers developed two methods to improve ideation. One approach casts HCI/UX in the role of a facilitator who gets data science teams and problem owners to brainstorm together~\cite{yildirim_creating_2023, yang2019unremarkable}. This work explicitly focused on helping innovators recognize situations where moderate model performance creates user value. It builds on the observation that making systems with moderate performance is much easier than making systems with excellent model performance. The other approach, AI Matchmaking~\cite{liu_human-centered_2024}, builds on a technology-centered innovation approach (matchmaking~\cite{bly_design_1999}). It centers user needs by including the problem owners in the matchmaking work. Both ideation approaches work to reduce the risks of technical viability and user acceptance. However, they do not address the financial or ethical risks.

The ideation phase typically involves two iterative steps: generating concepts and selecting the most promising ones. Bill Buxton emphasizes the importance of this selection process, aiming to ``discard more than we keep.~\cite{buxton_sketching_2007}'' He highlights that effective selection isn’t just about liking an idea, but making critical comparisons—asking ``Do I want this rather than that, and why?~\cite{buxton_sketching_2007}'' This approach underscores the necessity of evaluating concepts in parallel to determine their relative merits~\cite{dow_parallel_2011, tohidi_getting_2006}. However, recent efforts in early-stage AI innovation have primarily focused on idea generation, leaving a gap in our understanding of how to effectively select the concepts with the most potential.

\subsection{RAI Research on Mitigating Problems in Real-World Settings}
The RAI community has seen significant growth over the past decade with the increasing use of AI systems across various domains and a subsequent rise in AI harm cases. To address the ethical concerns arising from harmful AI systems, the RAI community has developed tools and processes that refine existing systems, document their limitations~\cite{mitchell_model_2019}, or facilitate audits and impact assessments~\cite{holstein_improving_2019, wong_seeing_2023}. However, there is a growing recognition that some ethical concerns are inherent to a specific problem formulation, requiring a fundamental system redesign rather than post hoc refinements \cite{raji_fallacy_2022, boyarskaya_overcoming_2020, holstein_improving_2019}. Some post hoc technical fixes such as “de-biasing” may inadvertently amplify the biases that they were supposed to address. In sum, much of the RAI work has focused on improving existing systems (i.e., making the thing right) rather than exploring which other AI concepts might have been more feasible in the first place (i.e., making the right thing). Recent studies investigating industry product teams' current practices and challenges around AI fairness, found that teams were most interested in finding ways to avoid ethical challenges in the first place.  

The RAI community has also developed databases and taxonomies that capture ongoing and emergent forms of algorithmic harm and may further help AI practitioners anticipate these potential harms~\cite{blodgett_language_2020, brundage_malicious_2024, noauthor_aiaaic_nodate}. However, AI systems are deployed in heterogeneous social contexts and interplay with social, cultural, and/or organizational dynamics~\cite{shelby2023sociotechnical}. This makes it difficult to anticipate harm. A design approach can help develop practical tools that are theoretically grounded in these taxonomies, center critical, actionable dimensions, and guide AI developers in systematically identifying potential harms before committing to system development. Moreover, the RAI community has predominantly focused on ethical risks. AI practitioners must ensure that AI concepts also create value for stakeholders. This requires a more holistic evaluation that incorporates concerns like financial and user-acceptance risks. 

Given these challenges, recent calls to action urge researchers to focus on the earliest phases of AI innovation, where teams can ideate more broadly (e.g.,~\cite{boyarskaya_overcoming_2020, holstein_improving_2019, raji_fallacy_2022, passi_problem_2019, wang_designing_2023}). The RAI community has shown a growing interest in leveraging HCI and design methodologies to integrate ethical considerations from the start~\cite{deng_understanding_2023, holstein_improving_2019, lee2024don}. Although some recent studies have examined these early stages~\cite{kawakami_situate_2024}, to our knowledge none have compared multiple concepts in parallel, and few have evaluated AI ideas beyond ethical concerns. This study contributes to the emerging body of work on RAI. It showcases how early-stage design probes can assist interdisciplinary teams in integrating RAI considerations with commercial goals. Teams can evaluate multiple AI concepts using a holistic and structured framework that weaves ethical concerns into commercial criteria of desirability, feasibility, and viability.

\section{Design Process Overview}
We were interested in the handoff between ideation and iteration when AI innovation teams assess many AI concepts and select what to make. We wanted to explore a proactive approach that enhances the coordination of interdisciplinary teams to identify low-risk, high-benefit concepts efficiently. We aim to integrate RAI considerations into the commercial development process to improve receptivity and acceptance of RAI in the private sector. This exploration seeks to understand how AI practitioners perceive and adapt to the envisioned workflow.

For this exploratory research, we chose a Research through Design (RtD) approach~\cite{zimmerman_research_2007}. RtD leverages the tools and processes of design practice to holistically explore problematic situations. It is useful for investigating new ways of designing. We conducted three interconnected design experiments. Design experiments are different from scientific experiments. Within the context of RtD research, “a design experiment includes [any] intentional actions design researchers take to further their understanding of the situation their work addresses”~\cite{zimmerman_recentering_2022}. Design Experiment 1 (DE1) explored examples from design and other fields of assessing and making a selection from many options. This informed and inspired the design of our probe materials. Design Experiment 2 (DE2) developed the materials for the probe study. This constructive activity operationalized insights from DE1. Design Experiment 3 (DE3) executed our probe study. It brought professional AI innovators together and had them enact a more formal approach to assessing and selecting an AI concept. 

HCI has a long history of using probes where researchers disrupt the current state in very intentional ways to observe participants’ reactions. They are meant to offer insights that inspire much more than to answer terse research questions, and they have been taken up and used in wildly different ways since their introduction~\cite{buchenau_experience_2000, gaver_design_1999, mattelmaki_design_2006, graham_how_2007, boehner_probes_2012}. 

Across the three design experiments, we employed a highly dialogical approach, continuously sharing and refining our findings through feedback from internal and external collaborators. Internally, our multidisciplinary team–experts in HCI, design, RAI, computer science, and psychology–engaged in brainstorming and design activities. Externally, we shared progress with two research groups: ~50 HCI researchers and ~30 RAI researchers. We used their feedback to inform the design of the probe and the questions we explored during the study with professionals.

\section{Design Experiment 1: Gathering Inspiration on Making Design Choices}

Designers often seek inspiration by broadly searching for relevant examples where others have addressed a similar situation~\cite{miller2014searching}. We built on this pattern by exploring examples of how design and other fields have approached the challenge of choosing one from many. We looked at examples from HCI, design, engineering, and business. We hoped to find ideas we could borrow in the design of our probe.

We began by brainstorming where to look. Next, we conducted a broad search across academic and non-academic sources. We started with the ACM Digital Library and Google Scholar, employing diverse keywords, including but not limited to ``early-stage concept selection,'' ``project selection,'' ``concept convergence,'' ``innovation evaluation instrument,'' ``sorting techniques,'' and ``ideation selection techniques.'' We next expanded our search to include widely adopted practices in the industry, identified through books, online media, and video demonstrations. We collectively reviewed the growing set of artifacts and methods using a process similar to critique where we talked about the usefulness of an example to our problematic situation. We repeatedly asked which examples supported consideration of technical, financial, user acceptance, and ethical risks and benefits for services, customers, users, and society.
        
Examples we drew inspiration from include:

\begin{itemize}
    \item Dot Voting: supports the collective intelligence of the group by allowing members to vote with ‘dots’ on various options~\cite{gibbons_dot_2019}.
    \item Impact-Effort Matrix: 2x2 matrix for a group to evaluate and prioritize tasks or projects based on the level of effort required and the impact they will have~\cite{gray_gamestorming_2010, gibbons_using_2018}.
    \item Morphological Charts: systematically generates new ideas and selects optimal solutions by decomposing a problem into its functions and listing various options for each, from which the best can be chosen~\cite{smith_concept_2012, van_boeijen_delft_2020}.
    \item Pugh Analysis: supports numeric rating and comparison of many possibilities by evaluating each option against a baseline using specific criteria, facilitating the identification of the most viable solution~\cite{pugh_concept_1981}.
    \item Harris Profile: visualizes and evaluates strengths and weaknesses of design concepts in relation to predefined design requirements~\cite{harris_new_1982, van_boeijen_delft_2020}.
    \item Risk Assessment Matrix: A prioritization framework (low-med-high) to decide which risk factors to prioritize, based on the severity and the likelihood of the risk happening~\cite{anthony_tonycox_jr_whats_2008}.
    \item Cooper’s New Product Selection Model: Thirteen factors that indicate the product’s potential for success, derived from 195 industrial product projects~\cite{cooper_empirically_1981}
\end{itemize}

Our critique of the examples led to the following key considerations for the probe design:
\begin{itemize}
    \item The process should strike a balance between having a formal structure and remaining designerly ~\cite{stolterman_nature_2008}. It should feel more like a practice method and less like a setup for a controlled study.
    \item Innovators should holistically consider the benefits and the four risks. They all interact with each other.
    \item The whole process must be fast. Teams should not spend more than 2 hours comparing 10 to 20 concepts to make a selection.
    \item The process should leverage the collective intelligence of the different disciplines and consideration of different organizational priorities.
    \item The process should compare multiple concepts in parallel, instead of focusing on one concept at a time~\cite{tohidi_getting_2006}.
\end{itemize}

\begin{figure*}
    \centering
    \includegraphics[width=\linewidth]{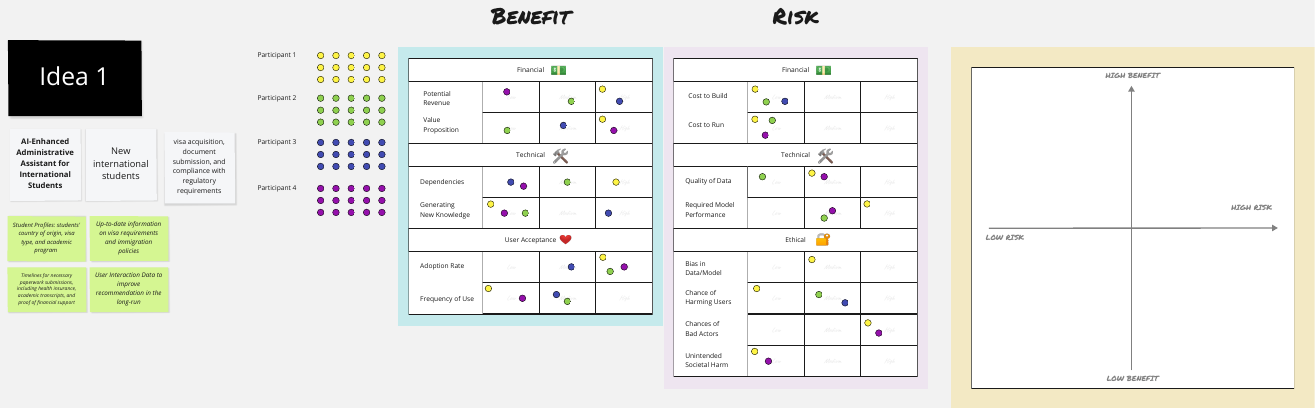}
    \caption{Initial prototype combining dot-voting with the 2x2 matrix approach.}
    \label{fig:initial-prototype}
    \Description{Sketch of a design probe used in an early-stage evaluation. On the left, a design concept is described using post-it notes; in the center, participants engaged in dot-voting on various parameters and questions; on the right, a 2×2 matrix maps perceived benefits and risks.}
\end{figure*}

\section{Design Experiment 2: Designing a Probe to Explore AI Concept Sorting and Responsible AI Integration}
Building on the criteria from DE1, we focused on making a probe that was flexible and could be used for rapid assessment. We designed a probe to provide professionals with scaffolding for assessing and rating a set of AI concepts. The scaffolding was meant to guide them toward high-benefit, low-risk concepts and away from high-risk, low-benefit. With respect to the holistic experience of the probe, we took inspiration from experience prototyping~\cite{buchenau_experience_2000}, particularly how it delivers critical aspects of a possible future experience and then asks participants to reflect on what they really want.

\subsection{Process Overview}
The process comprised three main activities: (i) brainstorming probe designs, (ii) assessing impacts, and (iii) refining structure and flow. We conducted iterative pilot testing to enhance our design process, including 13 formal pilot tests simulating the envisioned probe process and numerous informal tests. Feedback from diverse academic researchers and practitioners further enriched our approach.

\subsubsection{Brainstorming Probe Designs.} The team held five brainstorming sessions, where researchers sketched probe designs based on Design Experiment 1 criteria and refined them through group feedback. Broader team input helped prioritize and shape the ideas.

\subsubsection{Assessing Impacts.} To select a direction, we evaluated each concept against key criteria: (i) speed and usability in early-stage design, (ii) effective scaffolding of discussions across four areas (financial, technical, user acceptance, and ethical), and (iii) the ability to surface agreements and disagreements within teams. After extensive discussions, we chose an approach that combined dot-voting and a 2x2 matrix. This was intended to allow team members to categorize concepts into high-benefit, low-risk or low-benefit, high-risk groups and to facilitate focused discussions on prioritizing the concepts they believe to be most viable through voting. This framework also provided a simple, intuitive three-level rating system (high-medium-low) to assess each concept’s desirability, feasibility, and viability. Figure~\ref{fig:initial-prototype} shows the initial prototype.~\footnote{In the DE3 workshop study, we modified the probe's design to improve the online facilitation process by replacing dot-voting with a drop-down button format. Although the form changed, the content and functionality of the probe stayed consistent throughout the paper.}

\begin{figure*}
    \centering
    \includegraphics[width=\linewidth]{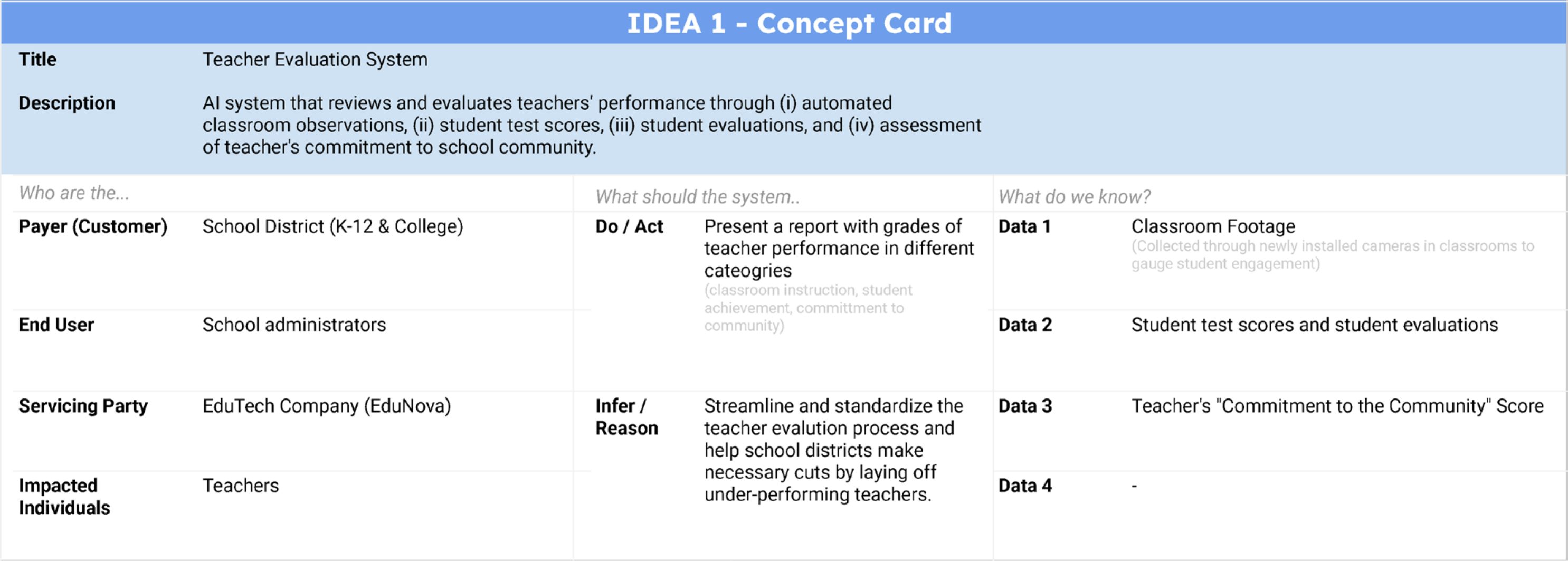}
    \caption{Concept Card with a top area for the title and description. The bottom area has three columns: stakeholders (left), system description (middle), and relevant datasets (right).}
    \label{fig:concept-card}
    \Description{Concept Card with a top area for the title and description. The bottom area has three columns: stakeholders (left), system description (middle), and relevant datasets (right).}
\end{figure*}

\begin{figure*}
    \centering
    \includegraphics[width=\linewidth]{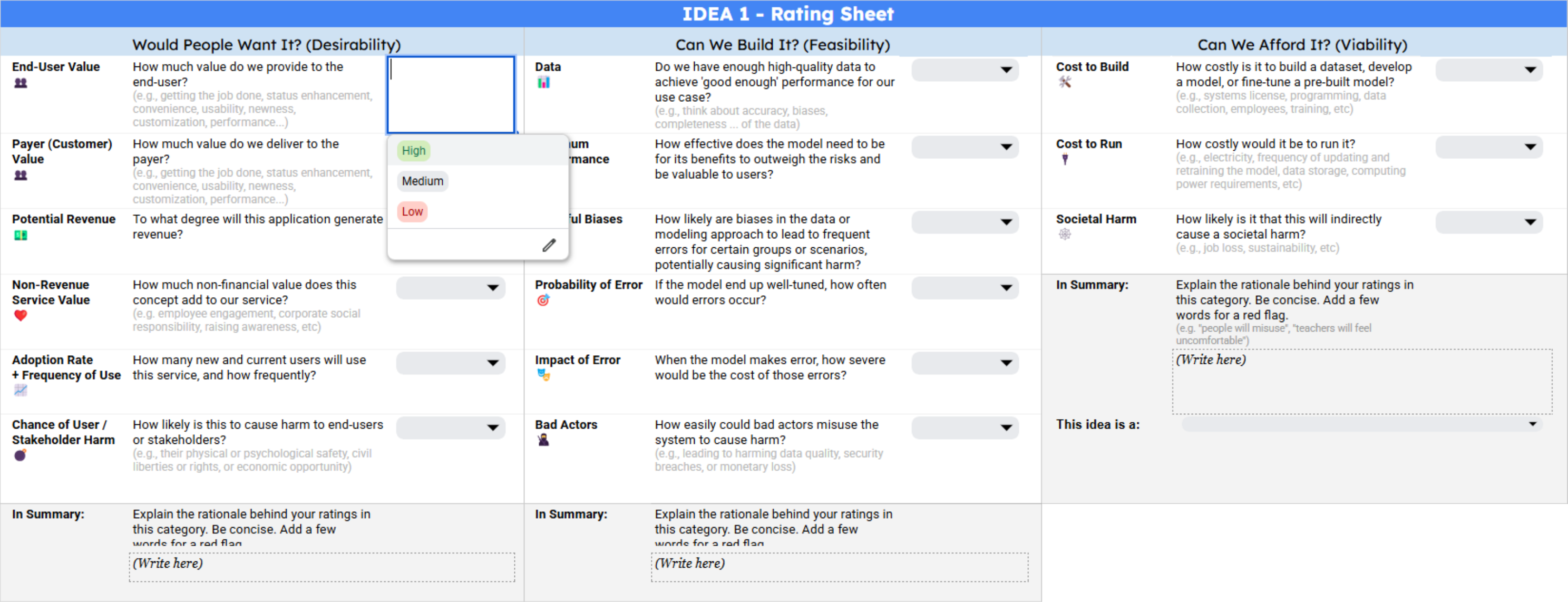}
    \caption{Individual Rating Sheet where participants rate each item as high, medium, or low using a dropdown menu. At the bottom of each column, they provide qualitative reasoning to support their quantitative ratings.}
    \label{fig:rating-sheet}
    \Description{Individual Rating Sheet where users rate each row as high, medium, or low using a toggle-down button. At the bottom of each column, users provide qualitative reasoning to support their quantitative ratings.}
\end{figure*}

\subsubsection{Refining the Form, Structure, and Flow.} Feedback from pilot tests with both students and professional industry teams continuously led to refinements in structure. In developing our prompt questions, we utilized DE1 inspirations for foundational guidance, selectively incorporating and synthesizing financial insights from business literature~\cite{henriksen1999practical, bitman2008conceptual, steele1994commercial, udell1982evaluating}, technical insights from classical input-model-output representations~\cite{googledeveloper_introduction_2022}, and both ethical and technical considerations from RAI literature~\cite{saxena_ai_2024, shelby2023sociotechnical, raji_fallacy_2022, wang_against_2024, kawakami_situate_2024, holstein_improving_2019}. Researchers then brainstormed prompts together. After multiple discussions between us, we organized them under desirability, feasibility, and viability. Inspired by prior research (e.g., ``Do-Reason-Know worksheet''~\cite{yildirim_sketching_2024}), we designed the Concept Card to outline the AI’s input-model-output mechanism. We also included stakeholder analysis, informed by repeated findings and feedback during the pilot tests, to address broader ecosystem concerns. Facilitation processes were streamlined through additional tests, reducing inefficiencies.

\begin{table*}[]
\caption{A list of prompting questions in an Individual Rating Sheet.}
\label{table:individual-rating-prompts}
\resizebox{0.85\textwidth}{!}{
\begin{tabular}{lll}
\toprule
\textbf{Category}                                                                                 & \textbf{Keyword}                                                             & \textbf{Prompt}                                                                                                                                                                                                 \\ \midrule
\multirow{6}{*}{\begin{tabular}[c]{@{}l@{}}Would People\\ Want It?\\ (Desirability)\end{tabular}} & \cellcolor{gray!15}End-User Value                                                               & \cellcolor{gray!15}\begin{tabular}[c]{@{}l@{}}How much value do we provide to the end-user? \\ (e.g., getting the job done, status enhancement, convenience, usability, newness, \\ customization, performance...)\end{tabular}  \vspace{0.03cm}  \\
                                                                                                  & \begin{tabular}[c]{@{}l@{}}Payer (Customer) \\ Value\end{tabular}            & \begin{tabular}[c]{@{}l@{}}How much value do we deliver to the payer? \\ (e.g., getting the job done, status enhancement, convenience, usability, newness, \\ customization, performance...)\end{tabular}  \vspace{0.03cm}     \\
                                                                                                  & \cellcolor{gray!15}Potential Revenue                                                            & \cellcolor{gray!15}To what degree will this application generate revenue?  \vspace{0.03cm}                                                                                                                                                        \\
                                                                                                  & \begin{tabular}[c]{@{}l@{}}Non-Revenue\\ Service Value\end{tabular}          & \begin{tabular}[c]{@{}l@{}}How much non-financial value does this concept add to our service? \\ (e.g. employee engagement, corporate social responsibility, raising awareness, etc)\end{tabular}   \vspace{0.03cm}            \\
                                                                                                  & \cellcolor{gray!15}\begin{tabular}[c]{@{}l@{}}Adoption Rate\\ + Frequency of Use\end{tabular}   & \cellcolor{gray!15}How many new and current users will use this service, and how frequently?   \vspace{0.03cm}                                                                                                                                    \\
                                                                                                  & \begin{tabular}[c]{@{}l@{}}Chance of User / \\ Stakeholder Harm\end{tabular} & \begin{tabular}[c]{@{}l@{}}How likely is this to cause harm to end-users or stakeholders? \\ (e.g., their physical or psychological safety, civil liberties or rights, or economic \\ opportunity)\end{tabular}\vspace{0.03cm} \\ \midrule
\multirow{6}{*}{\begin{tabular}[c]{@{}l@{}}Can We \\ Build It? \\ (Feasibility)\end{tabular}}     & \cellcolor{gray!15}Data                                                                         & \cellcolor{gray!15}\begin{tabular}[c]{@{}l@{}}Do we have enough high-quality data to achieve `good enough' performance for \\ our use case? \\ (e.g., think about accuracy, biases, completeness ... of the data)\end{tabular}   \vspace{0.03cm}  \\
                                                                                                  & \begin{tabular}[c]{@{}l@{}}Minimum\\ Performance\end{tabular}                & \begin{tabular}[c]{@{}l@{}}How effective does the model need to be for its benefits to outweigh the risks and \\ be valuable to users?\end{tabular} \vspace{0.03cm}                                                             \\
                                                                                                  & \cellcolor{gray!15} Harmful Biases                                                               & \cellcolor{gray!15}\begin{tabular}[c]{@{}l@{}}How likely are biases in the data or modeling approach to lead to frequent errors \\ for certain groups or scenarios, potentially causing significant harm?\end{tabular}\vspace{0.03cm}             \\
                                                                                                  & Probability of Error                                                         & If the model ends up well-tuned, how often would errors occur?\vspace{0.03cm}                                                                                                                                                  \\
                                                                                                  & \cellcolor{gray!15} Impact of Error                                                              & \cellcolor{gray!15}When the model makes errors, how severe would be the cost of those errors? \vspace{0.03cm}                                                                                                                                     \\
                                                                                                  & Bad Actors                                                                   & \begin{tabular}[c]{@{}l@{}}How easily could bad actors misuse the system to cause harm?\\ (e.g., leading to harming data quality, security breaches, or monetary loss)\end{tabular}\vspace{0.03cm}                             \\ \midrule
\multirow{3}{*}{\begin{tabular}[c]{@{}l@{}}Can We \\ Afford It? \\ (Viability)\end{tabular}}      & \cellcolor{gray!15} Cost to Build                                                                & \cellcolor{gray!15}\begin{tabular}[c]{@{}l@{}}How costly is it to build a dataset, develop a model, or fine-tune a pre-built model?\\ (e.g., systems license, programming, data collection, employees, training, etc)\end{tabular} \vspace{0.03cm}\\
                                                                                                  & Cost to Run                                                                  & \begin{tabular}[c]{@{}l@{}}How costly would it be to run it? \\ (e.g., electricity, frequency of updating and retraining the model, data storage, etc)\end{tabular}\vspace{0.03cm}                                             \\
                                                                                                  & \cellcolor{gray!15} Societal Harm                                                                & \cellcolor{gray!15}\begin{tabular}[c]{@{}l@{}}How likely is it that this will indirectly cause societal harm? \\ (e.g., job loss, sustainability, etc)\end{tabular}\vspace{0.03cm}                                                                \\ \bottomrule
\end{tabular}
}
\end{table*}

\subsection{Final Probe Design}
The final probe includes four artifacts and a workflow for utilizing them. Below, we describe the artifacts, instructions for using the artifacts, and outline the flow of the workshop session.

\subsubsection{Concept Card.}~\label{section:conceptcard} The concept card (Figure~\ref{fig:concept-card}) is a template completed prior to the workshop and includes four key components: (i) concept overview, (ii) stakeholder list, (iii) AI system description, and (iv) data. 

The concept overview provides a brief title and a short description summarizing the concept. The stakeholder list identifies four key roles: payer, end-user, servicing party, and impacted individual. The payer is the entity funding the service, while the end-user regularly interacts with the system. The servicing party distributes and manages the service. Lastly, the impacted individual includes those indirectly affected by the service, such as those experiencing unintended consequences. For example, in the case of Google Search, the end-user is the person searching for information, the payer is the advertisers, and impacted individuals might include website owners affected by search ranking algorithms. The system description adopts the ``Do-Reason-Know'' framework from Yildirim et al. (2024)~\cite{yildirim_sketching_2024}. It outlines (i) Do—the actions performed by the system (e.g., presenting recommended videos); (ii) Reason—the reasoning or inference process enabling the action (e.g., ranking videos by relevance while balancing diversity). Finally, the data column outlines (iii) Know—the computational data required to enable these actions (e.g., user data, metadata, or contextual information). 

The Concept Card format facilitates a uniform presentation of AI concepts, aiding those assessing multiple ideas. Pilot testing revealed that a consistent layout helps users quickly locate essential information, making comparisons easier and more efficient.

\subsubsection{Individual Rating Sheet}~\label{section:individualrating} The individual rating sheet (Figure~\ref{fig:rating-sheet}) helps people quickly assess AI innovation concepts based on four factors: financial, technical, user-acceptance, and ethical concerns. Designed to be completed in 5 minutes, it encourages informal, rapid evaluations using a High-Medium-Low scale paired with brief qualitative comments for context.

The sheet focuses on three main questions: (1) Desirability (“Wo-uld people want it?”), (2) Feasibility (“Can we build it?”), and (3) Viability (“Can we afford it?”), integrating ethical considerations into the framework. With 15 questions total (Table~\ref{table:individual-rating-prompts}), participants conclude by categorizing the concept as Keeper, Maybe, or Show-stopper.

\subsubsection{Team Response Overview}~\label{section:teamresponse}

\begin{figure*}
    \centering
    \includegraphics[width=\linewidth]{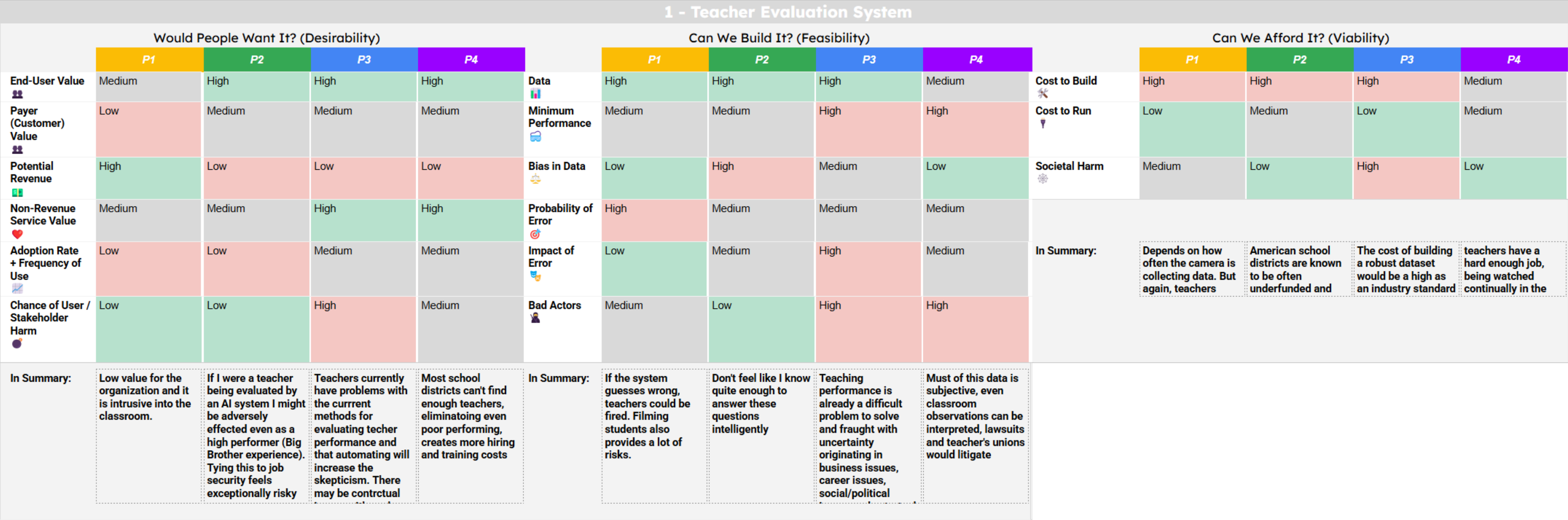}
    \caption{Team Response Overview consolidates individual responses from the rating sheet into a single view. Example shown is for illustrative purposes only and does not reflect actual ratings.}
    \label{fig:team-response-overview}
    \Description{Team Response Overview consolidates individual responses from the rating sheet into a single view. Example shown is for illustrative purposes only and does not reflect actual ratings.}
\end{figure*}

The Team Response Overview (Figure~\ref{fig:team-response-overview}) visualizes team ratings for a concept using color codes—green (positive), red (negative), and gray (neutral). Each column represents a question, with individual responses displayed side-by-side. Facilitators and team members use this visualization to quickly scan across each other’s perspectives as input for subsequent collective discussion. Here, the team can spot (dis)agreements, focusing discussions on critical issues for efficient and targeted decision-making.

\subsubsection{Risk-Benefit Matrix}~\label{section:riskbenefitmatrix}

\begin{figure}
    \centering
    \includegraphics[width=.9\linewidth]{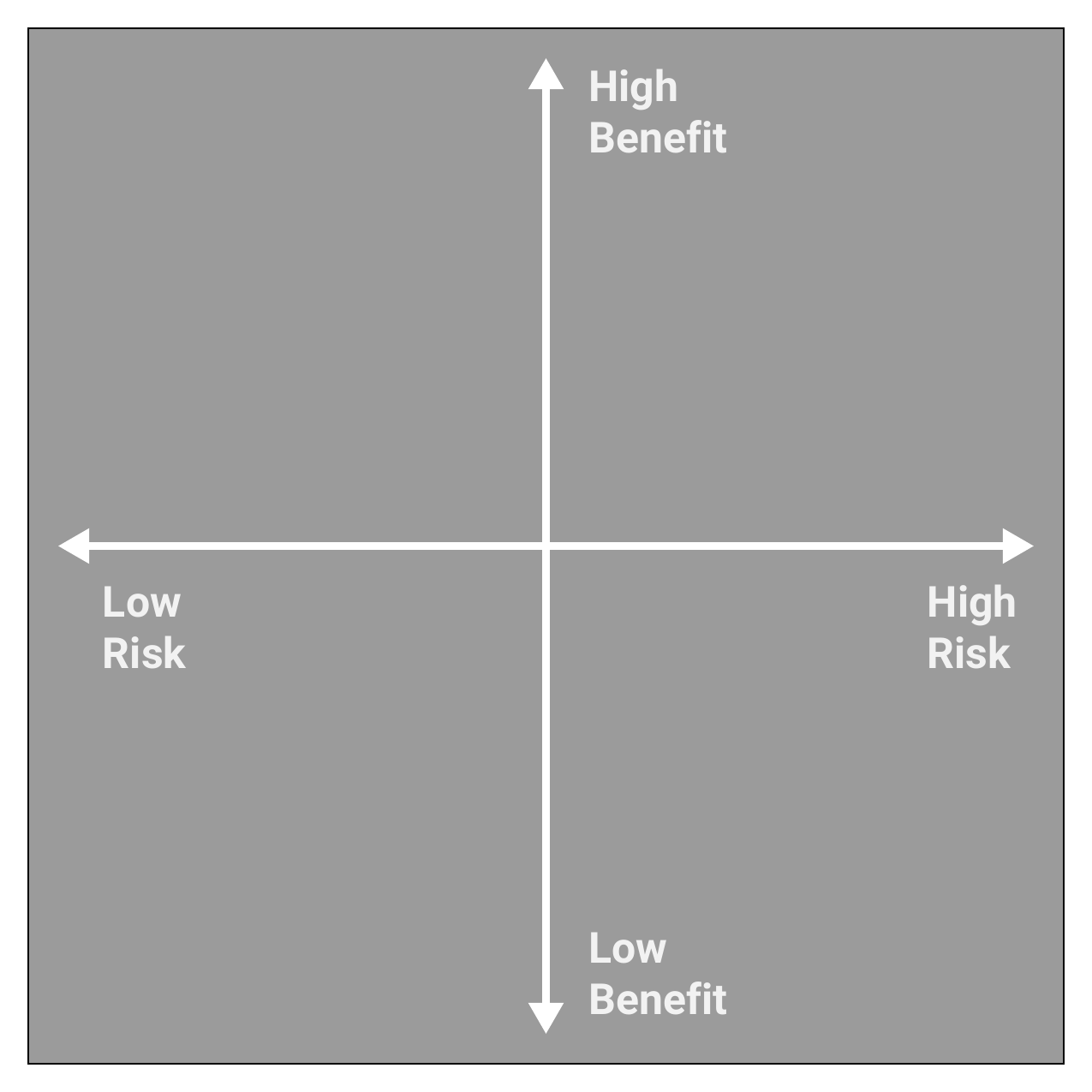}
    \caption{The Risk-Benefit Matrix, a 2x2 matrix with risk on the x-axis and benefit on the y-axis. The top-left quadrant marks the most desirable concepts, the bottom-right the least.}
    \label{fig:risk-benefit}
    \Description{The Risk-Benefit Matrix, a 2x2 matrix with risk on the x-axis and benefit on the y-axis. The top-left quadrant marks the most desirable concepts, the bottom-right the least.}
\end{figure}

The Risk-Benefit Matrix (Figure~\ref{fig:risk-benefit}) is a 2x2 matrix for evaluating concepts based on benefit (Y-axis) and risk (X-axis). Concepts in the top-left quadrant (high benefit, low risk) are prioritized, while those in the bottom-right (low benefit, high risk) are deprioritized or discarded. Quadrants in between prompt further discussion for refinement or strategic decision-making. Teams collaboratively place concepts on the matrix, with a facilitator guiding the process and referencing the Team Response Overview to address agreements and disagreements. 

The Risk-Benefit Matrix was inspired by the 2×2 structure of the Impact-Effort Matrix. In the Impact-Effort Matrix, effort represents the work required, and impact reflects the potential positive outcomes. While widely used in industry~\cite{gray_gamestorming_2010, gibbons_using_2018, admin_how_2022} and HCI research~\cite{yildirim_creating_2023, yildirim_investigating_2023}, it lacked the flexibility to address the broader range of risks and benefits needed for our project. For instance, ethical risks cannot be fully captured through the lens of effort. To address these limitations, we developed the Risk-Benefit Matrix as a more comprehensive framework.

\subsubsection{Workflow}

\begin{figure*}
    \centering
    \includegraphics[width=0.8\linewidth]{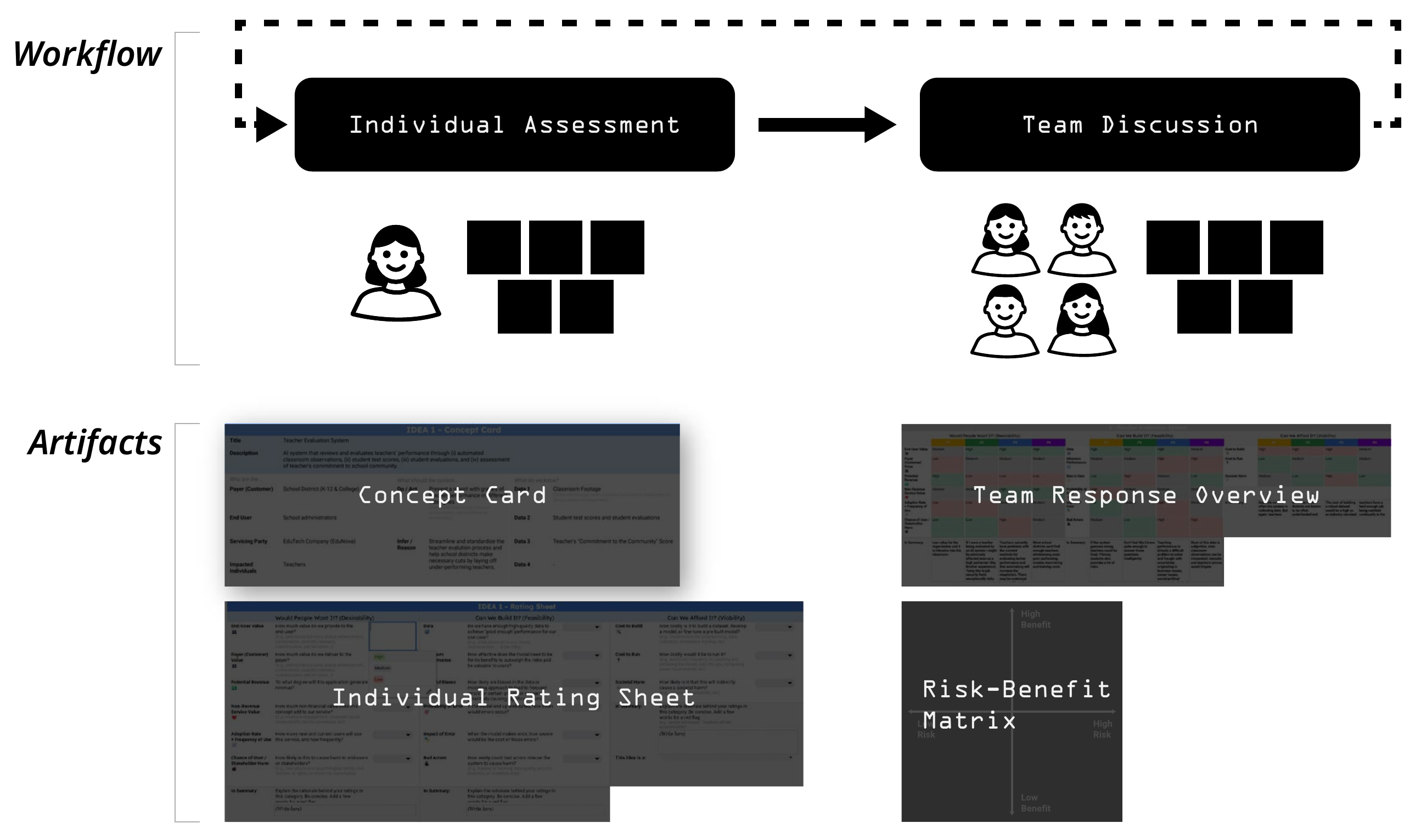}
    \caption{Visualized workflow of our probe. The top row illustrates the structure, beginning with individual assessments followed by team discussions on rated concepts. The bottom row highlights the artifacts used at each step.}
    \label{fig:workflow}
    \Description{Visualized workflow of our probe. The top row illustrates the structure, beginning with individual assessments followed by team discussions on rated concepts. The bottom row highlights the artifacts used at each step.}
\end{figure*}

Our probe’s workflow is a one-hour workshop designed for an interdisciplinary team, ideally including representatives covering financial, technical, user acceptance, and ethical concerns. The workshop focuses on assessing, sorting, and selecting AI concepts. Figure~\ref{fig:workflow} illustrates the overall workflow. Participants begin by reading the Concept Card (section~\ref{section:conceptcard}) and individually evaluating a subset of concepts using the Individual Rating Sheet (section~\ref{section:individualrating}). These individual ratings are then aggregated into a Team Response Overview (section~\ref{section:teamresponse}), which highlights areas of agreement and disagreement among participants. 

The facilitator plays a central role in managing the workshop, using the Team Response Overview to guide cross-domain discussions. They prioritize aspects with the most disagreements, encouraging dialogue to surface diverse perspectives and foster shared understanding. During team discussions, participants collaboratively map the concepts onto the Risk-Benefit Matrix (section~\ref{section:riskbenefitmatrix}), identifying high-benefit and low-risk concepts. This process is repeated for manageable subsets of concepts to maintain focus and ensure efficiency.

\subsection{Reflection}
Design Experiment 2 opened a window into the complexities of creating an early-stage AI design probe that could help teams sort and prioritize ideas with both confidence and care. Through this journey, we uncovered insights that shaped not only the probe itself but also the way we think about responsible innovation.

\subsubsection{Clarity in Concept Definition}
Early ideation thrives on creativity and abstraction, where ideas are fluid, bold, and unbound by constraints. Yet, when it came time to evaluate those ideas, participants struggled with vagueness. They needed something more concrete—details about technical dependencies, system actions, or clear definitions of stakeholders—to make informed decisions. This revealed a pivotal tension: how do you guide teams from the expansive freedom of brainstorming to the sharp clarity needed for meaningful evaluation? The Concept Card emerged as our bridge. We extensively explored what would be the bare minimum information to be able to rate the concept under four concerns.

\subsubsection{Integrating Ethical Concerns into Traditional Innovation}
As we developed our probe, we faced a key challenge: how to embed ethics into commercial innovation practice. Research revealed that practitioners often struggle to integrate ethics, citing reasons such as lack of guidance or institutional pressures, among others~\cite{prem_ethical_2023, rakova_where_2021, schiff_explaining_2021}. To address this, we wove ethics into the familiar framework of desirability, feasibility, and viability, making it a natural part of the evaluation process. This shift transformed ethics from a distant afterthought into a practical, integral component of innovation discussions, fostering more responsible and balanced decision-making.

\subsubsection{The Power of Facilitation}
Early pilot tests showed how critical the facilitator was in managing the interdisciplinary discussion. Without structure, discussions drifted, and participants spent far too long evaluating a single concept. By refining the facilitation process, we improved slow workshop progress (1 concept discussed for 1 hour) into a much more dynamic, collaborative experience. Skilled facilitators managed time effectively, keeping discussions focused while allowing room for debate. Our experience mirrors prior work that promotes the role of designers working as facilitators during the ideation of AI concepts~\cite{yildirim_how_2022, wang_designing_2023}.

\section{Design Experiment 3: Probe Study}
For DE3, we executed a probe study using the materials created in DE2. We wanted to explore resistance to and acceptance of a more formal assessment and selection process as well as the integration of RAI into what was meant to feel like a commercial development activity. We designed DE3 to help practitioners critically reflect on their future and present practices. 

\subsection{Study Design}
We recruited 15 AI innovation professionals using email, LinkedIn, and Slack. We divided participants into four interdisciplinary teams, with 3 to 4 on each team. During the study, participants individually rated and collectively sorted 6 early-stage AI concepts while we observed and took notes. Afterward, we interviewed them to gather insights about their experiences and how the process differed from their existing practices. The study design was submitted to and approved by our university’s Institutional Review Board (IRB).

%=================================================================
\noindent \textbf{\textit{Participants.}} We required all participants to be 18 years of age or older, and to live and work in the US. Recruitment focused on people with prior or current experience in AI innovation. We only accepted people with at least two years of experience innovating with AI. {Recruitment materials described the study as an opportunity to try a tool for early-stage AI concept assessment, with a focus on selecting high-benefit, low-risk ideas. The recruitment materials did not frame the study in terms of ethics or Responsible AI.} Participants came from diverse professional backgrounds, including design, data science, engineering, business strategy, and product management (see Table~\ref{table:participant} for details). Each received a \$30 gift card as compensation. All participants provided informed consent and signed consent forms prior to participation. 

\begin{table}[]
\caption{List of participants who attended the probe study in DE 3.}
\label{table:participant}
\resizebox{0.47\textwidth}{!}{
\begin{tabular}{llll}
\toprule
\textbf{ID} & \textbf{Team} & \textbf{\begin{tabular}[c]{@{}l@{}}Yrs of Professional \\ Experience\end{tabular}} & \textbf{Role}          \\ \midrule
\rowcolor{gray!15}P1                      & 1                & 6-10 yrs                                & UX Engineer            \\
P2                      & 1                & 2-5 yrs                                 & UX Designer            \\
\rowcolor{gray!15}P3                      & 1                & 21-30 yrs                               & UX Designer            \\
P4                      & 1                & 16-20 yrs                               & Product Manager        \\
\rowcolor{gray!15}P5                      & 2                & 11-15 yrs                               & Instructional Designer \\
P6                      & 2                & 11-15 yrs                               & Data Scientist         \\
\rowcolor{gray!15}P7                      & 2                & 31 yrs and above                        & Data Scientist         \\
P8                      & 2                & 2-5 yrs                                 & Software Engineer      \\
\rowcolor{gray!15}P9                      & 3                & 2-5 yrs                                 & UX Designer            \\
P10                     & 3                & 6-10 yrs                                & UX Designer            \\
\rowcolor{gray!15}P11                     & 3                & 2-5 yrs                                 & UX Designer            \\
P12                     & 3                & 6-10 yrs                                & Business Strategist    \\
\rowcolor{gray!15}P13                     & 4                & 6-10 yrs                                & Learning Engineer      \\
P14                     & 4                & 6-10 yrs                                & UX Designer            \\
\rowcolor{gray!15}P15                     & 4                & 11-15 yrs                               & Business Strategist    \\ \bottomrule
\end{tabular}
}
\end{table}

\begin{table*}[]
\caption{A list of concepts prepared prior to the workshop, with brief descriptions. Full details in the Concept Card format are provided in Appendix~\ref{section:appendix-concepts}.}
\label{table:concepts}
\resizebox{0.85\textwidth}{!}{
\begin{tabular}{lll}
\toprule
\# & \textbf{Title}                                                                                                       & \textbf{Description}                                                                                                                                                                                                                                                                                                                                                                                                    \\ \midrule
\rowcolor{gray!15}1  & \begin{tabular}[c]{@{}l@{}}Teacher Evaluation\\ System\end{tabular}                                                  & \begin{tabular}[c]{@{}l@{}}AI system that reviews and evaluates teachers' performance through \\ (i) automated classroom observations, (ii) student test scores, (iii) student \\ evaluations, and (iv) assessment of teacher's commitment to the school community.\end{tabular}\vspace{0.03cm}                                                                                                                                        \\
2  & \begin{tabular}[c]{@{}l@{}}AI-Driven Career\\ Counseling\end{tabular}                                                & \begin{tabular}[c]{@{}l@{}}A career advisory service that uses AI to analyze a student's skills, \\ interests, and job market trends to suggest suitable career paths and \\ necessary courses or skills development. It takes a special account of \\ future workforce forecasts.\end{tabular}\vspace{0.03cm}                                                                                                                         \\
\rowcolor{gray!15}3  & \begin{tabular}[c]{@{}l@{}}LLM-Based \\ Programming TA\end{tabular}                                                  & \begin{tabular}[c]{@{}l@{}}LLM-powered programming assistant in ``Introduction to Programming'' \\ courses that respond to students' conceptual questions about coding, \\ without revealing direct code solutions. It is expected to provide ongoing \\ and immediate support to students struggling with coding assignments, \\ just like a TA during office hours.\end{tabular}\vspace{0.03cm}                                        \\
4  & \begin{tabular}[c]{@{}l@{}}Test Scoring for Essays \&\\ Open-Ended Questions\\ for High School Students\end{tabular} & \begin{tabular}[c]{@{}l@{}}An AI system that automates the grading of essays and provides the \\ reason/rubric behind the grades and open-ended questions for high school \\ assignments and exams.\end{tabular}\vspace{0.03cm}                                                                                                                                                                                                        \\
\rowcolor{gray!15}5  & AI Proctor in Classrooms                                                                                             & \begin{tabular}[c]{@{}l@{}}During exams in the physical classroom, assess each student's behavior \\ by tracking their facial expressions, eye gaze, posture, and lip movement \\ in real-time and abnormal sounds in the room to flag the possibility of \\ cheating to the proctor. The video footage will be stored and accessible to \\ teachers for further review to determine if cheating occurred.\end{tabular}\vspace{0.03cm} \\
6  & AI-Powered Storyteller                                                                                               & \begin{tabular}[c]{@{}l@{}}An AI-driven application where middle-school kids can generate their own \\ stories, read them, and share them with classmates and friends. The tool \\ will read the story aloud and also generate illustrations for the stories. The \\ stories will be stored in the system to allow users to continually update them.\end{tabular}\vspace{0.03cm}                                                       \\ \bottomrule
\end{tabular}
}
\end{table*}

\begin{table*}[]
\caption{Workshop timetable that specifies the activities, time, and artifacts used in each step. }
\label{table:workshop-timetable}
\resizebox{0.85\textwidth}{!}{
\begin{tabular}{llll}
\toprule
\textbf{Activity}                                                     & \textbf{Artifact}                                                                      & \textbf{Duration} & \textbf{Description}                                                                                                                                                           \\ \midrule
\rowcolor{gray!15}Introduction                                                          & -                                                                                      & 20 min                                                               & \begin{tabular}[c]{@{}l@{}}Participants introduce themselves and build rapport \\ with their team. The facilitator outlines the workshop \\ goals and activities.\end{tabular}\vspace{0.03cm} \\
\begin{tabular}[c]{@{}l@{}}1st iteration - \\ Individual\end{tabular} & Individual Rating Sheet                                                                & 15 min                                                                & \begin{tabular}[c]{@{}l@{}}Participants individually rate three concepts using \\ the individual rating sheet.\end{tabular}\vspace{0.03cm}                                                    \\
\rowcolor{gray!15}\begin{tabular}[c]{@{}l@{}}1st iteration - \\ Team\end{tabular}       & \begin{tabular}[c]{@{}l@{}}Risk-Benefit Matrix, \\ Team Response Overview\end{tabular} & 10 min                                                                & \begin{tabular}[c]{@{}l@{}}The team collaboratively places each concept on \\ the Risk-Benefit Matrix, guided by the facilitator.\end{tabular}\vspace{0.03cm}                                 \\
\begin{tabular}[c]{@{}l@{}}2nd iteration - \\ Individual\end{tabular} & Individual Rating Sheet                                                                & 15 min                                                                & Repeat the 1st iteration process   \vspace{0.03cm}                                                                                                                                            \\
\rowcolor{gray!15}\begin{tabular}[c]{@{}l@{}}2nd iteration - \\ Team\end{tabular}       & \begin{tabular}[c]{@{}l@{}}Risk-Benefit Matrix,\\ Team Response Overview\end{tabular}  & 10 min                                                                & Repeat the 1st iteration process  \vspace{0.03cm}                                                                                                                                             \\
\begin{tabular}[c]{@{}l@{}}Post-workshop \\ Interview\end{tabular}    & -                                                                                      & 20 min                                                                & \begin{tabular}[c]{@{}l@{}}The facilitator asks participants reflective questions \\ about their experience.\end{tabular}\vspace{0.03cm}                                                      \\ \bottomrule
\end{tabular}
}
\end{table*}

%=================================================================
\noindent \textbf{\textit{Concepts.}} We prepared 6 AI concepts in the domain of education using our Concept Card structure. We selected concepts that showcase a range of benefits and risks, drawing inspiration from existing academic and industry examples, such as the AIAAC incident database for high-risk concepts. A summarized overview of the concept list is presented in Table~\ref{table:concepts}. The full details of the concepts are provided in Appendix~\ref{section:appendix-concepts}.

\begin{figure*}
    \centering
    \includegraphics[width=0.7\linewidth]{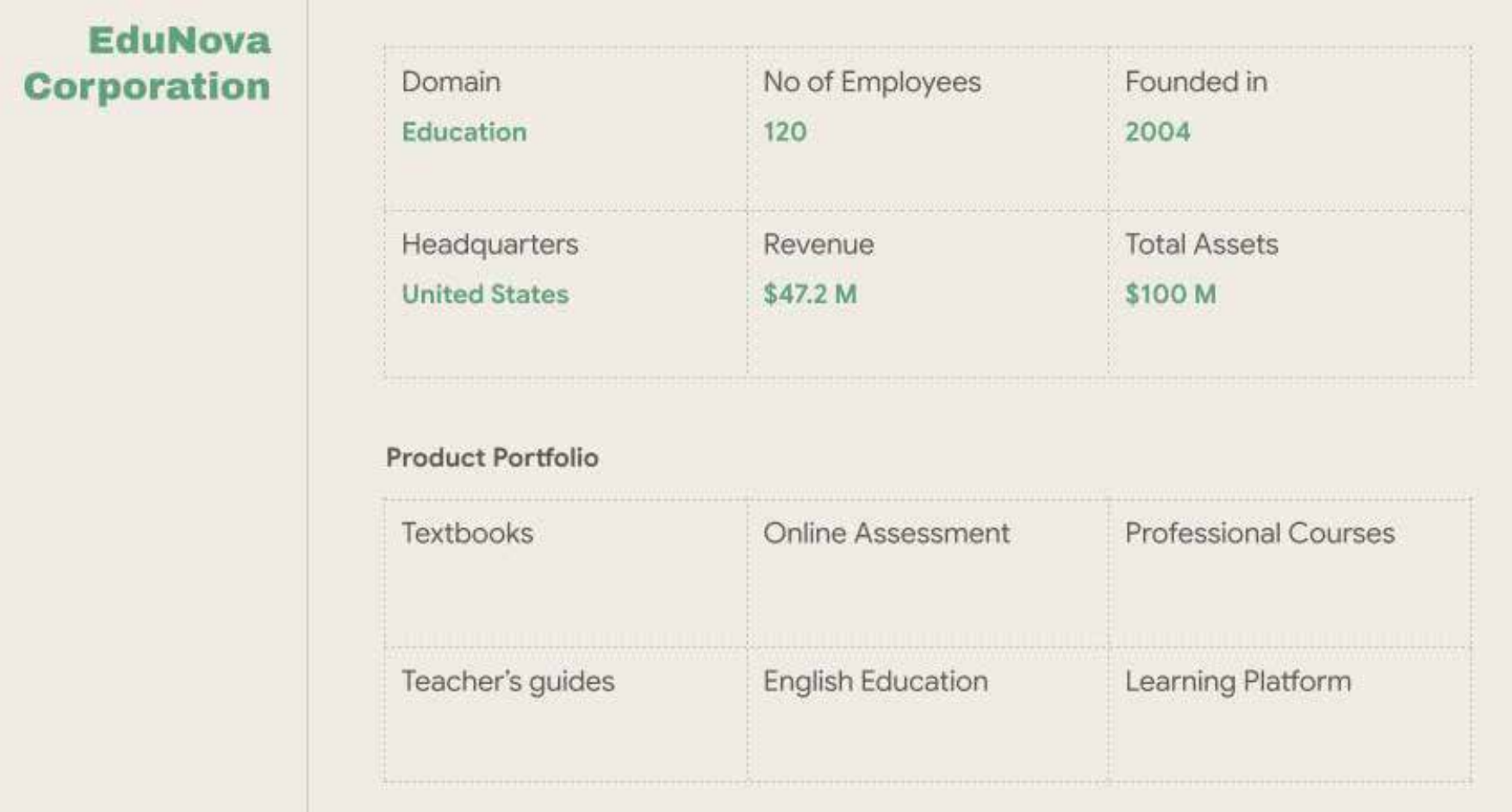}
    \caption{Fictional company profile that was shown to participants.}
    \label{fig:company-profile}
    \Description{Fictional company profile shown to participants, outlining a conceptual AI service provider. The profile includes sections on mission, services, data sources, and ethical commitments, designed to contextualize the proposed AI system during the design probe.}
\end{figure*}

%=================================================================
\noindent \textbf{\textit{Workshop.}} We ran the 90-minute workshops over Zoom. One researcher facilitated the sessions, while at least one other observed and took notes. The workshop structure was consistent across all teams (see Table~\ref{table:workshop-timetable} for the schedule). As a warm-up, we used a hypothetical scenario and asked participants to roleplay being coworkers at a fictional educational technology company. We provided an overview of the fictional company’s profile (see Figure~\ref{fig:company-profile}) to help establish context.

%=================================================================\
\noindent \textbf{\textit{Post-Workshop Interview.}} {We reserved the final 20 minutes of each session for a semi-structured, group-based interview conducted with all participants immediately following the design activity. These focus group-style discussions explored participants’ overall experience with assessing and selecting a concept, and how this process compared to their current practices. We also invited suggestions for improving this stage of AI innovation.} 
%We reserved the last 20 minutes for a semi-structured interview probing participants about their overall experience of assessing and selecting a concept and how this process differs from their current practice. We also probed for suggestions about how to improve this stage of AI innovation. 

%=================================================================
\noindent \textbf{\textit{Data Collection \& Analysis.}} We recorded and transcribed each workshop, and collected all artifacts, including individual rating sheets and the collective Risk-Benefit matrices. {For analysis, three researchers conducted affinity diagramming across transcripts, artifacts, and session notes to identify key themes and gain insights into how the design activities shaped participants' reflections and outcomes.}
%We analyzed the transcripts, artifacts, and our notes using affinity diagrams to identify key themes and derive insights.

%=================================================================
%=================================================================
%=================================================================
%=================================================================
%=================================================================
%=================================================================
\subsection{Findings}

\begin{figure*}
    \centering
    \includegraphics[width=0.95\linewidth]{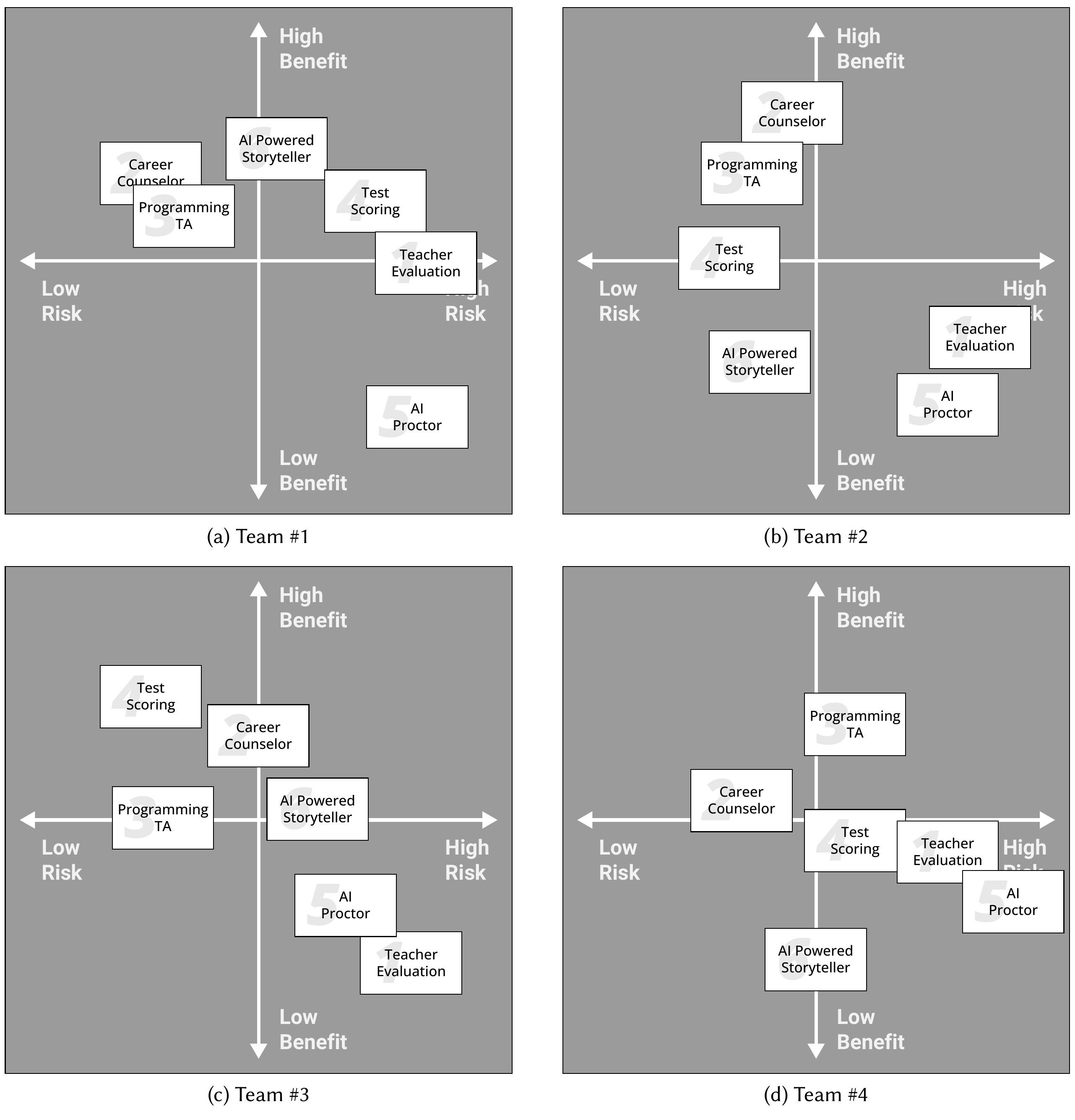}
    \caption{Risk-Benefit analysis of AI concepts across Teams 1–4. While all teams rated Concepts 1 and 5 as high-risk and low-benefit (bottom-right), Concepts 2 and 3 were often placed in the low-risk, high-benefit quadrant (top-left), with some teams placing other concepts there as well. }
    \label{fig:finding}
    \Description{Figure showing risk-benefit analysis of six AI concepts across four teams using 2×2 matrices. Concepts 1 (Teacher Evaluation) and 5 (AI Proctor) were consistently rated as high-risk and low-benefit. Concepts 2 (Career Counselor) and 3 (Programming TA) were frequently placed in the low-risk, high-benefit quadrant, though some variation exists across teams.}
\end{figure*}

All participants seemed eager to collaborate and to share their views. We found the team discussions to be dynamic and engaging. Disagreements about the risks and benefits were common; however, all teams consistently reached a consensus. As shown in Figure~\ref{fig:finding}, all teams placed Concepts \#1 and \#5 in the high-risk, low-benefit quadrant. (Note that although we developed our overall set of concepts to span a spectrum of risks and benefits, we did not have a specific ordering in mind.) They generally placed Concepts \#2 and \#3 in the low-risk, high-benefit quadrant. This provides some preliminary evidence that a more formal approach to assessing AI concepts can lead to consistent results across different teams. 

\subsubsection{Acceptance of RAI in Concept Selection}

Participants uniformly recognized RAI issues. The process allowed them to tie their RAI concerns to commercial goals and better justify why they considered an AI concept to be high-risk or low-benefit. %\textcolor{blue}{Notably, this receptiveness emerged even though the study was not presented as being focused on ethics or Responsible AI. Recruitment materials described the session as an opportunity to explore an AI innovation tool for identifying high-benefit, low-risk concepts.}
Far from being an afterthought, ethical concerns sometimes became the center of the discussion. For example, Concept \#1 (Teacher Evaluation) was rated high-risk by all four teams. They questioned its desirability, citing likely teacher opposition. They questioned its feasibility, noting potential harmful biases and a likely failure to meet performance standards. They also questioned its viability, arguing that the operational costs would surpass the benefits. While participants expected that the concept might offer medium to high-benefits to administrators (i.e. payer), all agreed it posed too great a risk to teachers, who were neither the end-users nor the payers. This led to an explicit and deliberate decision by participants to prioritize the ethical concerns for teachers, concluding that the ethical risks to impacted stakeholders far outweighed the potential financial benefits.

Ethical risks prompted rich, multidimensional discussions that considered intersections between  technical feasibility, user acceptance, ethical considerations, and financial viability. Discussing one dimension (e.g., technical feasibility) almost always prompted realizations in other dimensions (e.g., user acceptance or ethical considerations). For example, when evaluating Concept \#5 (AI Proctor), participants noted that collecting bio-privacy data to detect cheating behavior was unethical, technically infeasible, and likely expensive. While some participants recognized the potential high benefit of the system due to its scalability and human-in-the-loop design—which flags student behavior rather than imposing direct penalties—many raised concerns about its operation in a morally ambiguous area, such as defining what constitutes cheating. They highlighted the technical challenges in reliably detecting cheating, the difficulties in accurately labeling data affected by factors like `nervousness,' which can vary significantly due to a student's cultural background. Moreover, they stressed the high costs associated with errors in mislabeling students, which could severely damage the relationships between students and teachers, as well as between students and the school.

The Risk-Benefit Matrix facilitated nuanced discussions, emphasizing that the realization of potential benefits was contingent upon addressing identified risks. For example, in the case of Concept \#2 (Career Counseling), although it was positively assessed by all four teams, participants raised concerns about potential biases in job recommendations based on gender, race, and socioeconomic status present in historical data. They underscored that the concept's high potential benefits could only be achieved if these significant risks were effectively mitigated. This nuanced dialogue highlighted the critical need to resolve these issues before considering the implementation of the concept. 

This approach contrasts with the Impact-Effort matrix, which generally assumes the system is buildable and focuses primarily on the effort required for development. We noticed that discussions using the Risk-Benefit Matrix often involved questioning whether the application was even possible to build and examining the problem statement in a more fundamental manner. For instance, in discussing Concept \#4 (Test Scoring), participants examined how the system could potentially improve fairness and grading quality by flagging overlooked rubric components. However, they emphasized that such benefits were unrealizable if the potential technical risk of achieving consistency remained unaddressed.

Participants strongly emphasized the critical importance of considering policy and regulatory restrictions, such as HIPAA~\footnote{The Health Insurance Portability and Accountability Act (HIPAA)} and GDPR~\footnote{The General Data Protection Regulation (GDPR)}, early in the development of AI products. They highlighted these concerns as ``very real'' due to the additional costs and feasibility challenges they introduce. This entails making preliminary plans for data handling and addressing security issues, which they identified as key factors. Furthermore, they noted that in their practice, they not only assess the difficulty of building the system but also strive to quantify the extent of these challenges. They viewed policy risk as equally important to the four risks our probe explored, and they seemed almost puzzled that it was not included. Our probe seems to have effectively prompted participants to bring up these rich ideas and suggestions on early-stage considerations.

\subsubsection{Selection Behavior}

We observed that discussions about risks quickly led to consensus among the team, allowing for swift agreement about where a concept fit on the low to high-risk dimension of the Risk-Benefit matrix. However, reaching agreement on benefits was more challenging. It often sparked extensive discussions and careful evaluations of trade-offs among stakeholders. Participants frequently questioned, ``but benefits for whom?'' For example, during the discussion of Concept \#3 (Programming TA) there was significant debate about whether the university's potential financial gain justified possibly reducing learning opportunities for students to be trained as tutors, alongside affecting their income opportunities. Concept \#2 (Career Counseling) sparked debates on whether the benefits to end-users outweighed those to the universities themselves. With Concept \#4 (Test Scoring), participants struggled to assess benefits, recognizing that while teachers would significantly benefit, they were not the ones financing the service. Participants noted the need to exercise caution, recognizing that the benefits valued by school districts might not necessarily align with those of the teachers, as evidenced by the current system's failure to compensate teachers for unpaid grading tasks performed outside of school hours.

Participants often based initial judgments on the status quo when assessing the benefits and risks. However, the team discussion when using the matrix helped them to more deeply interrogate concepts. For example, in discussions about Concept \#2 (Career Counseling), some participants expressed skepticism about AI's effectiveness in offering career suggestions to students, citing their limited capabilities. Other participants expressed that current school career counselors often provide limited insights to students, despite the evident needs and justified school spending in this area. Some participants felt that Concept \#2 could provide `good enough' suggestions and had the potential to reach a broader range of students more effectively, potentially motivating students to reflect more on their goals. Similarly, the perceived benefits of Concept \#3 (Programming TA) were diminished by the existence of ChatGPT, as participants questioned the additional value it could provide to students who already use ChatGPT to seek help and complete their assignments.

We observed that participants frequently referenced information from the Concept Cards as they articulated and voiced their concerns. For example, in discussions about Concept \#1 (Teacher Evaluation), they highlighted the subjectivity of listed data components (i.e., classroom observation, teacher commitment) and the impossibility of establishing a definitive measure of teacher performance, questioning the validity of the outcome itself, which is a key ethical consideration~\cite{jacobs2021measurment}. Similarly, for Concept \#5 (AI Proctor), participants highlighted the `video footage' data component, expressing concerns about the financial burden of data collection and training. Additionally, they raised issues regarding the complexity and privacy concerns of managing biodata, questioning the value of such efforts given the minimal benefits. In discussing Concept \#2 (Career Counseling), participants mentioned concerns about the costs associated with consistently updating the model with rapidly-changing job market data. They also noted the large volume of data and the significant effort required to train and fine-tune the model to ensure accuracy and relevance.
 
Participants noted that the collective ranking and discussion activities were critical for evaluating concepts. Collective discussion allowed team members to highlight strengths and weaknesses together. Observing agreements and disagreements helped participants better understand different perspectives, often arising from their disciplinary perspectives. They all seemed open to changing their minds about issues that are outside of their area of expertise. For example, many shared how initial uncertainties were resolved after hearing their teammates articulate their reasoning, prompting them to adjust their initial assessments.

Participants felt that assessing concepts using individual Rating Sheets before proceeding to collective discussions gave them clarity and confidence in forming their opinions. However, they pointed out that individual ratings would have limited use without contrasting them with others’ viewpoints. When assessing the concept alone, participants recognized they were making numerous judgments and often lacked confidence in these. Through team dialogue, these assumptions were clarified, leading to a consensus and more confident judgment about the concept. Additionally, the prompts from Individual Rating Sheet encouraged critical thinking and stepping outside comfort zones, which enhanced their understanding of the concepts. This preparatory phase ensured that collective discussions were informed and productive.

Unexpectedly, participants demonstrated a desire to refine and reframe concepts, rather than simply selecting among the existing concepts as-is. They often redescribed concepts in ways that increased benefits and lowered risks. For instance, during discussions about Concept \#4 (Test Scoring), one participant suggested changing the customer and reframing the concept as an AP prep tool to increase its appeal to payers. Another proposed positioning it as a teacher companion tool, offering example essays for comparison rather than direct scoring. In almost all cases, they made changes to the business aspects, not the technical aspects of the concept.

\subsubsection{Reflections on Current Practices}

Majority of the participants shared that they did not participate in deciding what new products or services their organization should make. In reflecting on their practice, most noted that they did participate in decision-making for incremental improvements to a current application. Participants from larger organizations also shared that introducing a new product or service impacts existing product ecosystems, a risk our probe did not address. Almost all felt our probe pushed them into doing work and making decisions outside of their normal responsibilities. Several participants noted that they participated in design sprints as part of an Agile development process. They shared that sprints could lead to a pivot, a change to a new application or a new target market. This serial process is similar to our probe, but it lacks the ideation of many possible concepts proceeding assessment and selection. 

The few participants who said they were actively involved in concept selection (e.g., product managers) talked about following a structured process by using tools such as RICE (Reach, Impact, Confidence, and Effort) scoring and Impact-Effort matrices. They described decision-making on AI innovation within their organizations as hierarchical. They noted that “final” decisions were often overridden by board members or investors, who prioritized ambitious, high-risk concepts driven by market trends. They explained that senior executives and the tech culture often dismissed concerns about feasibility risks, believing that all technical challenges could be resolved with sheer effort. These statements align with prior work suggesting managers and executives overestimate benefits and underestimate the technical risks associated with AI innovation. This sentiment was shared across all four workshop teams. Some participants described it as engaging in ``watercooler shit talk.'' They shared that they frequently complained and questioned their colleagues about the real value of the AI concept that they were working on. Others expressed a feeling that the corporation pushes for AI merely for the sake of it. %A few noted that even after their team puts in a lot of effort to find a great business case, it will sometimes be overruled by senior executives.

They shared that decision-making varied significantly based on organization size and their title. Participants from larger organizations described a top-down approach, where key decisions were made at senior levels. Smaller organizations employed decision-making that heavily prioritized technical feasibility and speed. Participants from startups, for example, shared that their decisions were driven by the need to quickly develop a Minimum Viable Product (MVP), often within just a few days. Concepts that could not meet short timelines were deprioritized or ignored.

Participants noted differences between their current practices and the workflow our probe suggested. They described their approach as more `iterative,' involving building MVPs and pivoting based on feedback. They noted that in their current practice, they did not typically examine all risks and benefits simultaneously or comprehensively. They describe the issues of technical feasibility, financial viability, user acceptance, and ethical risks as showing up individually, at different stages of the development process. The fast pace and earlier commitment to model development is consistent with previous studies suggesting there might be a lack of ideation in AI innovation~\cite{yildirim_investigating_2023, yildirim_creating_2023}. The lack of rigorous ideation and focused attention on MVPs and development partially explains why there are many AI innovation tools and methods for prototyping and almost nothing for ideation.

\subsubsection{Concept Card}

When asked what aspects of the probe stood out as particularly useful, participants unexpectedly highlighted the Concept Card. Initially, we viewed it as a secondary part of our probe, assuming it was tangential to the overall process. However, participants emphasized that its structured template was instrumental in providing the necessary information for sound reasoning about the concepts. Some even expressed interest in adopting the concept card for their own organizations.

Participants noted that the concept card filled a significant gap in their current decision-making processes. They particularly liked how it clarified feasibility by listing relevant data and outlining what the system could realistically achieve. This level of detail not only grounded their evaluations, but also ensured that discussions focused on concepts presented with comparable degrees of specificity.

The differentiation between stakeholders—end-users, payers, and impacted stakeholders—was also highlighted. Participants found this helpful for identifying the root causes of a concept's challenges. For instance, in discussions about Concept \#4 (Test Scoring), participants noted it was beneficial for end-users (teachers) but less so for payers (schools) and impacted stakeholders (students). The clarity around stakeholders enriched the conversations, enabling teams to assess benefits and risks with greater precision and insight.

\section{Discussion}
Our study explored a more structured approach to considering risks and benefits at the point of project selection. Through three design experiments, we investigated how RAI considerations could be integrated into commercial innovation practices at an early stage and how the process might scaffold teams to avoid pursuing problematic concepts that could lead to failure. These experiments highlighted key gaps in current workflows and provided actionable insights for improving future methods.

Below, we delve into the challenges and opportunities associated with executing early-stage ideation and selection processes in AI innovation. We explore the implications of integrating RAI into commercial frameworks and examine how service design can aid this integration. Additionally, we reflect on the implications of our HCI research methods, identify areas for improvement, and propose open questions for further research.

\subsection{Challenges and Opportunities of Ideation in AI Innovation}

Our findings suggest that the AI innovation process might be negatively dominated by top-down management decision-making. Our participants did not regularly participate in brainstorming new applications or in the selection of what to make. They only participated in the ideation of small, incremental improvements. Their reactions raise questions. Who is doing ideation? Is anyone doing the ideation? Is AI innovation driven by one-off ideas from managers? Design and HCI research shows the importance of ideation to develop more effective and successful products and services~\cite{dow_parallel_2011}. Furthermore, previous studies have shown that people tend to give higher ratings to a design when presented with a single option and are more hesitant to criticize it, compared to when they are shown multiple concepts simultaneously ~\cite{tohidi_getting_2006}. Yet we saw little evidence that ideation was happening. Prior HCI suggested a lack of ideation for AI innovation might be due to the fact that data scientists were not trained to ideate~\cite{yildirim_sketching_2024}; however, our probe hints that organizational leaders don’t support innovation teams in ideating.

Our observation that managers might be envisioning and selecting concepts raises additional concerns about a problematic innovation process. Previous research shows low AI literacy among executives with only 3\% of executives at S\&P 500 companies viewed as AI literate~\cite{benlian_why_2024}. Prior RAI work raised this concern, noting that ethical assessments in AI are primarily considered by senior management~\cite{ayling_putting_2022}. It seems like the people who know the least about AI’s technical challenges and about how it might create unintended harm are playing a very large role in determining the best way to get value from AI’s opportunity. If this really is the current state of the industry, it begins to explain the tension participants shared about effectively communicating technical risks to executives with decision-making power. 

Despite being largely excluded from decision-making processes around what to build, our interdisciplinary teams showed strong capabilities for doing this work and doing it in a way that tapped into their collective intelligence. Participants engaged in thoughtful discussions about risks and benefits, achieving consistency across teams in promoting and demoting concepts based on benefits and risks. Their frustrations when talking about their leaders indicate their desire to participate in ideation and selection. With both the desire and the skills, the question now becomes one of changing organizational culture to unleash the benefit of their participation. Future research should study the implications and potential benefits of actively including product-facing roles in the decision-making process. This includes examining current workflows and practices to identify who is truly driving decision-making and where ideation originates. By exploring these dynamics, we can move towards a deeper understanding of AI innovation.

Our research reveals a potential gap in how HCI and UX design are taught and the needs and processes in the industry. Practitioners working at Youtube have spoken about the need to differentiate three types of user-centered innovation, calling these “Versioning, Visioning, and Venturing”~\cite{dame_power_2019, conversions_product_2019, mannheim_experimenting_2021}. Versioning involves small, incremental improvements to a current product. Visioning focuses on creating a new product or major new feature and has a time scale running to three years. Our probe of assessing and selecting concepts fit visioning, not the versioning activities our participants were familiar with. We don’t see this distinction of three types of innovation as particularly new. Versioning, venturing, and visioning seem like an almost perfect mapping to McKinsey’s “three horizons framework”~\cite{mckinsey_quarterly_enduring_2009, blank_mckinseys_2019}. Interestingly, in HCI research and education, we seem to talk about innovation and create methods for doing HCI practice as if these three different types of making did not exist. We sort of teach a single approach to user-centered design, ignoring that the scale of the innovation under consideration matters. It might be time to advance our own research and curriculums to better fit what seem to be very real distinctions being used in practice.

\subsection{Supporting Early-Stage RAI Practices in Industry}

In response to the high failure rates of AI projects, several technology and consulting firms have started using centralized AI governance teams as one way of addressing risks~\cite{PwC_AIFactory, IBM_Garage, KPMG_Ignition, Deloitte_AI_Institute}. These teams (e.g., PwC AI Factory~\cite{PwC_AIFactory}) are generally staffed with AI practitioners and domain experts who decide which AI concepts the company should pursue. They are tasked with identifying and assessing AI use cases, staffing AI projects, adopting RAI practices, ensuring compliance with regulation, and supporting the development and deployment processes; thereby creating a one-stop-shop for the governance of AI products and services. 

These innovation teams suggest adopting RAI practices from the start, however, research in RAI shows that there is insufficient support in operationalizing RAI principles in practice~\cite{holstein_improving_2019, madaio_co-designing_2020, rakova_where_2021, kawakami_situate_2024}. RAI tools have supported innovation practice in various ways (e.g., data and model documentation~\cite{gebru_datasheets_2021, mitchell_model_2019, crisan_interactive_2022}, detecting and mitigation bias~\cite{bellamy_ai_2018, hardt_amazon_2021}, etc.) but there is a misalignment. RAI work has often framed ethical risks as standalone considerations that must be addressed on top of the practitioners’ usual tasks. Consequently, ethical concerns are often sidestepped or deprioritized due to tight project timelines and organizational pressures~\cite{rakova_where_2021}.   

In practice, innovation teams must assess multiple AI concepts together and scaffold this assessment within commercial goals. Prior RAI research suggests that ethical concerns often surface informally and are vulnerable to project constraints~\cite{rakova_where_2021}. Yet, our findings reveal three key insights for RAI – 1) ethical risks are intertwined with commercial goals and impact the desirability, feasibility, and viability of AI concepts, and 2) embedding ethical considerations into commercial criteria can prompt more nuanced discussions and holistic evaluations. Far from being sidelined, ethical risks often took precedence during discussions, with some concepts being rejected primarily due to their ethical implications, 3) assessing multiple concepts together allowed the practitioners to identify high-risk AI concepts as well as lower-risk AI concepts that could be refined further.

This suggests that structured tools can elevate ethical considerations to a central role in decision-making without displacing other business priorities. This integrated approach can help innovation teams discover critical flaws before resources are invested, systematically avoid early missteps, and prioritize higher-value and lower-risk concepts.

\subsection{Centering Responsible AI Work in Service Design}

Our approach embraces more of a service design as opposed to a UX design perspective. We look at both the financial aspects and the larger ecology of stakeholders involved and impacted by the AI system~\cite{shelby2023sociotechnical}. We grounded our tools in service design to emphasize a holistic view of the entire service ecosystem, including stakeholders, organizational resources, and processes.

Prior work in RAI shows how ethical concerns are often overshadowed by commercial goals~\cite{rakova_where_2021, holstein_improving_2019}. However, by focusing on the broader ecosystem that will be impacted by an AI concept, we wove ethical questions directly into business-focused discussions, allowing participants to see ethical concerns as a complimentary priority and not a competing one. For example, Concept \#1 (Teacher Evaluation), even though financially appealing, was deemed too risky due to potential harm to educators. This more holistic framing shows how balancing commercial and ethical factors from the beginning can foster more nuanced conversations and inform responsible decisions.

The concept card explicitly maps out payers, end-users, and impacted stakeholders, a core service design principle that asks designers to differentiate between different user groups and consider each group’s unique needs, risks, and benefits. This breakdown appeared to have a huge impact on participants’ ability to understand a concept, compare it with other concepts, and reason about its risks and benefits. Discussing varying benefits and risks based on different stakeholders almost felt natural for them. We observed that deliberately listing it out for the assessment process helped facilitate and make this process of negotiating trade-offs more transparent. The key sticking point revolved around who should be receiving the most benefits and how this might make a concept better or worse.  

Iterative learning and a co-creative environment are central to a service design process. Our participants shared their expertise when comparing AI concepts side-by-side and identified shared challenges and risks. This ability to compare multiple concepts simultaneously fostered richer discussions and subsequently led to the participants’ seeking to refine AI concepts to lower their risk. Participants proposed actionable refinements to weaker concepts, such as repositioning the Test Scoring tool to better serve teachers rather than automating grading. Comparative evaluation, supported by tools like the Risk-Benefit Matrix, can encourage more nuanced decision-making and creative problem-solving.

By focusing on the broader AI ecosystem and not just technical factors, our probe materials collectively functioned as a boundary object, facilitating effective cross-disciplinary collaboration. Participants from diverse roles—data scientists, designers, and domain experts—were able to contribute effectively to discussions by identifying ethical risks that others might have missed. These tools ensured that all voices were heard and helped bridge gaps in knowledge and expertise.

HCI does not hold a strong concern for the financial aspects of innovation. Most methods attend to user needs, not to the financial needs of service providers. HCI research rarely explores the financial aspects of design concepts, which are critical to a product or service's commercial success. Our paper contributes to HCI research by integrating this perspective, offering a more holistic view of the potential impact a design concept can have on society. Our integrations of financial, technical, user acceptance, and ethical considerations offer some evidence that HCI, RAI, and business considerations and priorities can effectively be brought together.

\subsection{Lessons from Piloting with Students vs. Professionals}
We conducted pilot tests, largely with students, and then ran probe sessions with professionals. Their strikingly different behaviors, concerns, and reactions raise issues with respect to how our research community will often use students as a proxy for professionals. While students provided valuable perspectives, they often struggled with unfamiliar workflows and hesitated to confidently share clear opinions. Their assessments were often very slow, and we had concerns that professionals would not be able to quickly assess the concepts. In contrast, the professional teams adopted the probe with ease, leveraging their industry experience to evaluate concepts quickly and effectively while considering market and organizational factors. This contrast highlighted the critical importance of seeking feedback from real-world audiences—such as AI innovation teams—for generating meaningful and actionable insights in HCI studies.

\subsection{Limitations}
One limitation of our study is that participants evaluated concepts they had not envisioned. We realize that sorting behavior may differ when participants evaluate concepts they have ideated themselves. We made this choice due to time constraints and the challenge of recruiting enough professional AI innovators who all work in the same domain. Future research should explore how self-generated concepts may affect sorting outcomes and team dynamics. Furthermore, our study involved ad-hoc teams with no prior rapport, and who had little if any expertise in educational technology. While participants quickly adapted to the process and showed little struggle in evaluating the AI concepts, future studies should examine how the probe may perform within established teams in real organizational settings, where shared histories and domain expertise may influence results. {Additionally, as noted in the Discussion, our probe focused on visioning-scale concepts; had we used more versioning-oriented concepts with smaller variances, participants’ engagement with the tool might have differed.} Finally, the probe was designed for use within a single organization, but its applicability to multi-stakeholder collaborations or cross-organizational projects remains untested. Future work should investigate how the tool can be scaled or adapted for broader, more complex contexts.

\section{Conclusion}
This paper addresses a critical challenge in early-stage AI innovation: how to effectively ideate, sort, and select AI concepts while integrating RAI considerations into the process. By designing and deploying a probe, we demonstrated how interdisciplinary teams can navigate complex trade-offs, integrating RAI considerations seamlessly with technical, financial, and user-acceptance factors. Our findings suggest that we might have an overly top-down innovation process, coupled with a lack of thorough ideation that restricts full exploration of potential solution space. This work bridges the gap between RAI principles and real-world practice, offers the HCI and design community an empirical insight for embedding RAI concerns into AI systems, and suggests a mechanism for participation from a broader number of stakeholders on a team. By advancing our understanding of early-stage AI innovation practices, we position design as a driving force in developing AI systems that are more beneficial and less risky in terms of financial, technical, user acceptance, and ethical considerations.

%=================================================
%=================================================

\begin{acks}
    This research is supported by the National Science Foundation under Grant No. (2007501), the Presidential Postdoctoral Fellowship Program (PPFP) at Carnegie Mellon University, the Digital Transformation and Innovation Center at Carnegie Mellon University sponsored by PwC, the UL Research Institutes through the Center for Advancing Safety of Machine Intelligence, and the Block Center for Technology and Society at Carnegie Mellon University. Any opinions, findings, conclusions, or recommendations expressed in this material are those of the authors and do not necessarily reflect the views of our sponsors or research partners. We would also like to thank Dan Saffer, Raelin Musuraca, Laura Vinchesi, and Skip Shelly for their valuable feedback, as well as our anonymous reviewers for their insightful comments and suggestions.
\end{acks}

\bibliographystyle{ACM-Reference-Format}
\bibliography{references}

\newpage
\appendix

\section{Study Material: 6 AI Concepts}~\label{section:appendix-concepts}

\subsection{Concept 1: Teacher Evaluation System}
\begin{itemize}
    %\item \textbf{Title:} Teacher Evaluation System
    \item \textbf{Description:} AI system that reviews and evaluates teachers' performance through (i) automated classroom observations, (ii) student test scores, (iii) student evaluations, and (iv) assessment of teacher's commitment to school community.
    \item \textbf{Stakeholder} (\textit{Who are the...})
    \begin{itemize}
        \item \textbf{Payer:} School District (K-12 \& College)
        \item \textbf{End-User:} School administrators
        \item \textbf{Servicing Party:} EduTech Company (EduNova)
        \item \textbf{Impacted Stakeholder:} Teachers
    \end{itemize}
    \item \textbf{System description} (\textit{What should the system...})
    \begin{itemize}
        \item \textbf{Do/Act:} Present a report with grades of teacher performance in different categories (classroom instruction, student achievement, commitment to community)
        \item \textbf{Infer/Reason:} Streamline and standardize the teacher evaluation process and help school districts make necessary cuts by laying off under-performing teachers.
    \end{itemize}
    \item \textbf{Datasets} (\textit{What do we know?})
    \begin{itemize}
        \item \textbf{Data 1:} Classroom Footage (Collected through newly installed cameras in classrooms to gauge student engagement)
        \item \textbf{Data 2:} Student test scores and student evaluations
        \item \textbf{Data 3:} Teacher's "Commitment to the Community" Score
    \end{itemize}
\end{itemize}

\subsection{Concept 2: AI-Driven Career Counselling}
\begin{itemize}
    %\item \textbf{Title:} AI-Driven Career Counselling
    \item \textbf{Description:} A career advisory service that uses AI to analyze a student's skills, interests, and job market trends to suggest suitable career paths and necessary courses or skills development. It takes a special account of future workforce forecasts.
    \item \textbf{Stakeholder} (\textit{Who are the...})
    \begin{itemize}
        \item \textbf{Payer:} Highschool Parents, College Students
        \item \textbf{End-User:} High School and College Students
        \item \textbf{Servicing Party:} EduTech Company (EduNova)
        \item \textbf{Impacted Stakeholder:} Students
    \end{itemize}
    \item \textbf{System description} (\textit{What should the system...})
    \begin{itemize}
        \item \textbf{Do/Act:} Present a list of potential career paths using student academic records, extra-curricular activities, and interests and explain why the jobs were recommended to students.
        \item \textbf{Infer/Reason:} From the datasets, find a good career match for the student.
    \end{itemize}
    \item \textbf{Datasets} (\textit{What do we know?})
    \begin{itemize}
        \item \textbf{Data 1:} Job Market Data (current job listings and descriptions, previous job market trends and tendencies, industry trends and future job market predictions).
        \item \textbf{Data 2:} Educational Data (Course catalogs from various educational institutions)
        \item \textbf{Data 3:} Student Data. (1) Academic records and transcripts. (2) Extracurricular activities and achievements. (3) Personality assessments and career interest surveys.
        \item \textbf{Data 4:} Alumni Data. (Career paths of graduates from different programs)
    \end{itemize}
\end{itemize}

\subsection{Concept 3: LLM-based Programming TA}
\begin{itemize}
    %\item \textbf{Title:} LLM-based Programming TA
    \item \textbf{Description:} LLM-powered programming assistant in "Introduction to Programming" courses that respond to students' conceptual questions about coding, without revealing direct code solutions. It is expected to provide ongoing and immediate support to students struggling with coding assignments, just like a TA during office hours. 
    \item \textbf{Stakeholder} (\textit{Who are the...})
    \begin{itemize}
        \item \textbf{Payer:} Universities
        \item \textbf{End-User:} College students in "Introduction to Programming" course
        \item \textbf{Servicing Party:} EduTech Company (EduNova)
    \end{itemize}
    \item \textbf{System description} (\textit{What should the system...})
    \begin{itemize}
        \item \textbf{Do/Act:} Generate conversational answers to students' technical questions
        \item \textbf{Infer/Reason:} Understand students' questions and generate conversational replies without revealing direct code solutions. It tells students when you have arrived at the correct answer.
    \end{itemize}
    \item \textbf{Datasets} (\textit{What do we know?})
    \begin{itemize}
        \item \textbf{Data 1:} Few-Shot Learning Examples. (Current job listings and descriptions, Industry trends and future job market predictions)
        \item \textbf{Data 2:} Programming Language Document (i.e., Python, javascript, etc)
        \item \textbf{Data 3:} Code Execution Traces. (Data on how code is executed, including variable values, function calls, and control flow.)
    \end{itemize}
\end{itemize}

\subsection{Concept 4: Test Scoring for \& Open-ended Essay Questions for High School Students}
\begin{itemize}
    %\item \textbf{Title: } Test Scoring for Essays & Open-ended Questions for High School Students
    \item \textbf{Description:} An AI system that automates the grading of essays and provides the reason/rubric behind the grades and open-ended questions for highschool assignments and exams.
    \item \textbf{Stakeholder} (\textit{Who are the...})
    \begin{itemize}
        \item \textbf{Payer:} School District (High School)
        \item \textbf{End-User:} Teachers
        \item \textbf{Servicing Party:} EduTech Company (EduNova)
    \end{itemize}
    \item \textbf{System description} (\textit{What should the system...})
    \begin{itemize}
        \item \textbf{Do/Act:} Grade the quality of student responses and provide percentage points.
        \item \textbf{Infer/Reason:} Compare student answers to the right / "high-quality" answers.
    \end{itemize}
    \item \textbf{Datasets} (\textit{What do we know?})
    \begin{itemize}
        \item \textbf{Data 1:} Sample Answers and Essays (Annotated datasets of sample answers for open-ended questions, including different levels of quality and correctness.)
        \item \textbf{Data 2:} Scoring Rubrics (Rubrics and guidelines for grading open-ended questions and essays, specifying criteria for different levels of performance.)
    \end{itemize}
\end{itemize}

\subsection{Concept 5: AI Proctor in Classrooms}
\begin{itemize}
    %\item \textbf{Title:} AI Proctor in Classrooms
    \item \textbf{Description:} During exams in the physical classroom, assess each student's behavior by tracking their facial expressions, eye gaze, posture, and lip movement in real-time and abnormal sounds in the room to flag the possibility of cheating to the proctor. The video footage will be stored and accessible to teachers for further review to determine if cheating occurred.
    \item \textbf{Stakeholder} (\textit{Who are the...})
    \begin{itemize}
        \item \textbf{Payer:} School District (K-12 \& College)
        \item \textbf{End-User:} High School and College Students
        \item \textbf{Servicing Party:} EduTech Company (EduNova)
        \item \textbf{Impacted Stakeholder:} Students
    \end{itemize}
    \item \textbf{System description} (\textit{What should the system...})
    \begin{itemize}
        \item \textbf{Do/Act:} Flag suspicious cheating behavior by analyzing each students' behavior
        \item \textbf{Infer/Reason:} Identify suspicious cheating behavior by analyzing real-time video and audio (i.e., eye gaze, facial expressions, posture, movement of lips, abnormal sounds in the room).
    \end{itemize}
    \item \textbf{Datasets} (\textit{What do we know?})
    \begin{itemize}
        \item \textbf{Data 1:} Video Footage of Exams (Large amounts of labeled video footage from classroom exams showing both typical and suspicious behaviors).
        \item \textbf{Data 2:} Facial Expressions (Dataset of various facial expressions and head movements during exams, labeled as normal or suspicious).
        \item \textbf{Data 3:} Body Language (Annotated data of different body postures and movements, such as turning their heads often, which might indicate cheating).
        \item \textbf{Data 4:} Audio Recordings of Exams (Labeled audio data from exam settings, identifying normal sounds (e.g., rustling of papers) and suspicious sounds (e.g., whispering)).
    \end{itemize}
\end{itemize}

\subsection{Concept 6: AI-Powered Storyteller}
\begin{itemize}
    %\item \textbf{Title:} AI-Powered Storyteller
    \item \textbf{Description:} An AI-driven application where middle-school kids can generate their own stories, read them, and share them with classmates and friends. The tool will read the story aloud and also generate illustrations for the stories. The stories will be stored in the system to allow users to continually update them.
    \item \textbf{Stakeholder} (\textit{Who are the...})
    \begin{itemize}
        \item \textbf{Payer:} School District (Middle School)
        \item \textbf{End-User:} Middle School Students
        \item \textbf{Servicing Party:} EduTech Company (EduNova)
    \end{itemize}
    \item \textbf{System description} (\textit{What should the system...})
    \begin{itemize}
        \item \textbf{Do/Act:} Generate young adult (middle school level) stories. ; Generate illustrations for stories. ; Read these stories in different voices.
        \item \textbf{Infer/Reason:} Generate stories based on characters and plots provided by students. Update the story based on new information (e.g., plot twists).
    \end{itemize}
    \item \textbf{Datasets} (\textit{What do we know?})
    \begin{itemize}
        \item \textbf{Data 1:} Young-Adult(middle school) Story to fine-tune LLM (A large dataset of stories, including fairy tales, fables, and modern literature, to fine-tune an LLM like GPT-3).
        \item \textbf{Data 2:} Story Templates (Datasets of various story templates and outlines that can serve as the foundation for story generation).
        \item \textbf{Data 3:} Vocabulary Lists (Age-appropriate vocabulary lists to ensure the generated stories are suitable for the target audience).
        \item \textbf{Data 4:} Voice Data (Audio recordings of stories being read aloud, including different voices and intonations, to train text-to-speech models).
    \end{itemize}
\end{itemize}

\end{document}